\documentclass[useAMS,usenatbib,onecolumn]{mn2e}
\usepackage{graphicx}

\title[Effective power-law dependence of Lyapunov exponents on the central 
mass in galaxies]
{Effective power-law dependence of Lyapunov exponents on the central mass 
in galaxies}
\author[N. Delis, C. Efthymiopoulos and C. Kalapotharakos]
{
N. Delis$^{1,2}$
\thanks{E-mail: delnikle@gmail.com},
C. Efthymiopoulos$^{2}$
\thanks{E-mail: cefthim@academyofathens.gr} and
C. Kalapotharakos$^{3}$
\thanks{E-mail: constantinos.kalapotharakos@nasa.gov}\\
$^{1}$Department of Physics, University of Athens, 
Panepistimiopolis, 115 21 Athens, Greece\\
$^{2}$Research Center for Astronomy and Applied Mathematics, 
Academy of Athens, Soranou Efessiou 4, 115 27 Athens, Greece\\
$^{3}$University of Maryland, College Park (UMDCP/CRESST), College Park, 
MD 20742, USA, and\\
Astrophysics Science Division, NASA/Goddard Space Flight 
Center, Greenbelt, MD20771, USA
}
\begin{document}

\date{}

\pagerange{\pageref{firstpage}--\pageref{lastpage}} \pubyear{2002}

\maketitle

\label{firstpage}

\begin{abstract}
Using both numerical and analytical approaches, we demonstrate the existence
of an effective power-law relation $L\propto m^p$ between the mean Lyapunov 
exponent $L$ of stellar orbits chaotically scattered by a supermassive black 
hole in the center of a galaxy and the mass parameter $m$, i.e. ratio of the 
mass of the black hole over the mass of the galaxy. The exponent $p$ is found 
numerically to obtain values in the range $p \approx 0.3$--$0.5$. We propose 
a theoretical interpretation of these exponents, based on estimates of local 
`stretching numbers', i.e. local Lyapunov exponents at successive transits 
of the orbits through the black hole's sphere of influence. We thus predict 
$p=2/3-q$ with $q\approx 0.1$--$0.2$.  Our basic model refers to elliptical 
galaxy models with a central core. However, we find numerically that an 
effective power law scaling of $L$ with $m$ holds also in models with 
central cusp, beyond a mass scale up to which chaos is dominated by the 
influence of the cusp itself. We finally show numerically that an analogous 
law exists also in disc galaxies with rotating bars. In the latter case, 
chaotic scattering by the black hole affects mainly populations of thick 
tube-like orbits surrounding some low-order branches of the $x_1$ family 
of periodic orbits, as well as its bifurcations at low-order resonances, 
mainly the Inner Lindbland resonance and the 4/1 resonance. Implications 
of the correlations between $L$ and $m$ to determining the rate of 
secular evolution of galaxies are discussed. 
\end{abstract}

\begin{keywords}
galaxies -- dynamics; galaxies -- chaos; galaxies -- central black holes
\end{keywords}

\section{Introduction}  
There is by now overwhelming evidence that black holes with masses
ranging from $10^6$ to  $10^{10} M_\odot$ (see review \cite{ferrford2005}) 
exist in the center of most galaxies \cite{kormrich1995}, 
\cite{gebhardtetal1996}, \cite{faberetal1997}, \cite{kormendy1997}, 
\cite{kormendy1998}, \cite{vandermaletal1997}, \cite{gerbhardtetal2000}, 
\cite{gultetala2009}, \cite{gultetalb2009}, \cite{kormendy2013}, 
\cite{mcconma2013}. The presence of a supermassive 
black hole leads to a number of consequences for dynamics, whose study has 
been a subject of extended research in the last three decades 
(indicative references closely related to our present work are 
\cite{gerhbinn1985}, \cite{merrfrid1996}, \cite{mervalu1996},  
\cite{fridmanmer1997}, \cite{kandrsid2002}, \cite{kalvogcon2004}, 
\cite{kalvog2005}, see Merritt (1999) or (2013) for complete 
reviews of the subject.)

Inter alia, the growth of a central black hole provides a mechanism 
driving \emph{secular evolution} in galaxies. The creation of secularly 
evolving models by the insertion of a central mass is ubiquitous in N-body 
simulations of non rotating elliptical galaxies (\cite{merrquinl1998}, 
\cite{holbocketal2001}, \cite{holbocketal2002}, \cite{convogcal2002}, 
\cite{kalvog2005}, \cite{muzzcarpwach2005}, \cite{jessnaabburk2005}, 
\cite{muzzio2006}, \cite{kal2008}, \cite{valur2010}, \cite{vasath2012}). 
The secular evolution causes two main effects: 
i) the density profile becomes more cuspy in the center (\cite{holbocketal2002}), 
and ii) the galaxy becomes more spherical in the center and more axisymmetric 
in the outer parts (\cite{merrquinl1998}, \cite{kalvogcon2004}, \cite{kalvog2005}). 
On the other hand, the growth of black holes in disc galaxies slowly disrupts 
the bars by changing the stability properties of many orbits that support 
the bar (\cite{pfenn1984}, \cite{pfenndezeu1989}, \cite{hasetal1993}, 
\cite{normshellhas1996}, \cite{shen2004} (see review \cite{Debat2006}). 

In the case of elliptical galaxies, a physical interpretation of why,  
after the insertion of a central mass, secular evolution favors particular 
endstates was provided in \cite{kalvogcon2004} and \cite{kalvog2005} 
(see also \cite{efthvogkal2007}, \cite{kal2008}). This is based on closely 
examining the orbital dynamics of those particles whose orbits are directly 
affected by the central mass. The main scenario of the secular evolution 
process presented in these works can be summarized as follows: at a first 
stage, the insertion (or gradual 'turning on') of the central mass results 
in a conversion of the majority of box orbits into chaotic orbits 
(\cite{gerhbinn1985}). Following the evolution of the system by N-body 
simulations, it is found that the newly formed chaotic orbits start 
inducing a gradual change in the distribution of matter, i.e., the shape 
of the system, which becomes more spherical given that the distribution 
of chaotic orbits is more isotropic. The slow change in shape, in turn, 
causes an adiabatic change in the potential, thus affecting the 
{\it phase space structure}, in particular at energies for which the 
phase space was initially (before the insertion of the central mass) 
occupied mainly by box orbits passing arbitrarily close to the center. 
The main change of the phase space structure regards the volume 
increase of the domain corresponding to regular short axis tube (SAT) 
orbits (see \cite{binntre2008} for definition). As the volume of the SAT 
domain increases, some chaotic orbits are adiabatically captured inside the 
boundary of the SAT domain (\cite{kalvogcon2004}). This capture then induces 
an additional change in the form of the system, enhancing the conversion 
of chaotic orbits into SAT orbits, etc. This cyclic process maintains 
secular evolution up to a point where the population of chaotic orbits 
decreases substantially.  The systems evolved by this mechanism are closer 
to oblate. Furthermore, in the final stages the percentage of chaotic 
orbits becomes smaller even than the one in the original systems, 
before the insertion of the central mass. 

It should be noted that the degree up to which the transformation 
of a system from triaxial to axisymmetric proceeds depends on how many 
box orbits are transformed to chaotically scattered orbits due to the 
central mass. For example, in \cite{holbocketal2002} the initial percentage 
of outer box orbits is such that an adiabatic introduction of the central 
mass does not destroy triaxiality in the outer parts of their galaxy 
simulation models. Also, \cite{valur2010} examined how stochastic 
(or `non-reversible') can the whole process of secular evolution be 
characterized, by considering the secular evolution of dark matter 
haloes in various models of central mass concentrations. While their 
findings re-confirm the picture of (non-reversible) stochastic 
evolution in the case of a point-like central mass concentration 
(e.g. a black hole), they find that there is also a different, i.e., 
`regular' (or reversible) type of evolution taking place in systems in 
which the central mass is spatially distributed (e.g. a galactic disc 
or bulge embedded in a triaxial halo).

Hereafter, we focus on the mechanism of secular evolution induced by 
the chaotic scattering of centrophylic orbits. In order to quantify 
the rate of secular evolution induced by the above mechanism, 
\cite{kal2008} introduced a novel quantity, measurable in N-body 
simulations, called the \textit{effective chaotic momentum}
\begin{equation}\label{efcm}
{\cal L}={{{\Delta}N_w\over N_{total}} L_w}~~.
\end{equation}
In Eq.(\ref{efcm}), $L_w $ is the mean value of the {\it Lyapunov 
Characteristic Number} (LCN) of a sub-ensemble of chaotic orbits in the system 
after the insertion of the central mass. This is defined by the orbits belonging 
to a percentage window $\pm 0.3$ around the characteristic value where the 
distribution $P(\log L_j)$ of the logarithms of the Lyapunov exponents $L_j$ 
of all the particles in chaotic orbits presents its global maximum. 
Considering a percentage window is necessary since, as we will see also 
below, the distribution $P(\log L_j)$ is two-peaked, while the secular 
evolution is caused mainly by orbits forming the higher of the two peaks 
of the distribution. As a rule, these are centrophilic orbits passing 
arbitrarily closely to the central mass. On the other hand, ${\Delta}N_w$ 
is the total mass inside the same window, while $N_{total}$ is the total 
mass of the N-body system considered. 

The use of the effective chaotic momentum ${\cal L}$ allows one 
to quantify several phenomena related to the rate of secular evolution. 
A remarkable result is that there appears to be a global (i.e. the same 
in all simulations) theshold value ${\cal L}_{th}$ such that 
as a system undergoes secular evolution, with a time-varying value 
of ${\cal L}(t)$, the evolution becomes inefficient over a Hubble time 
when ${\cal L}(t)$ falls below ${\cal L}_{th}$ (\cite{kal2008}). 

From the two factors in the definition of ${\cal L}$ (Eq.\ref{efcm}), 
the percentage of chaotic orbits $\Delta N_w/N_{total}$ depends on 
the specific system studied, i.e. on the initial distribution function 
that determines the initial conditions of the simulation. However, 
as noted in \cite{kalvogcon2004} and \cite{kal2008}, the distribution 
of Lyapunov exponents, and in particular the value of $L_w$ is found 
numerically to be not very sensitive to the choice of initial distribution 
function in the simulations. Thus, simulations representing systems with 
different profiles and triaxiality were found to exhibit different 
percentages of chaotic orbits, but similar levels of $L_w$, the latter 
found, instead, to be correlated with the {\it value of the 
mass ratio} of the central mass over the mass of the hosting system. 
Hereafter, this ratio is simply referred to as the `central mass 
parameter' $m$. These findings indicate that, from the two factors entering 
the computation of the rate of secular evolution via the `effective chaotic 
momentum', one (percentage $\Delta N_w/N_{total}$ of chaotic centrophylic 
orbits) depends on the self-consistent distribution function of the system 
considered, while the other (mean Lyapunov exponent $L_w$) depends 
strongly only on the value of the mass parameter $m$.

In the present paper we focus on this latter dependence, and seek 
to model its dynamical origin. We note that a dependence of the mean 
Lyapunov exponent $L$ of the centrophilic orbits on $m$ is a result 
derived also in studies using fixed potentials (e.g. 
\cite{gerhbinn1985}, \cite{mervalu1996}, \cite{kandrsid2002}). 
In \cite{kal2008}, on the other hand, this dependence was explicitly 
determined by the orbital data of the particles in N-body simulations  
yielding the value of $L_w$ which is a good measure of $L$. A remarkably 
good {\it power-law dependence} was found, namely
\begin{equation}\label{lm}
L \propto m^p,~~~p\approx 0.5~~~.
\end{equation}
The proximity of $p$ to 0.5 was also found in models of simple galactic 
potentials \cite{kal2008}. As shown in section 2, somewhat smaller values, 
around $p\approx 0.3$, are deduced by a careful a posteriori analysis 
of the numerical results presented in \cite{mervalu1996} and in 
\cite{kandrsid2002}.

In the present paper we first reconfirm numerically the power-law 
(\ref{lm}) in further fixed-potential computations. Then, we develop a 
theoretical modelling allowing to interpret the origin of this power-law. 
We also justify why the exponent $p$ obtains values in the observed range. 
In particular, our theory yields an exponent $p=2/3-q$, where $q\sim 
0.1$--$0.2$. 

Our theoretical modeling stems from considering the dynamics of chaotic 
centrophilic orbits which undergo `transits', i.e. they spend part of 
their time inside and another part outside the so-called `sphere of influence' 
of the central mass (\cite{gerhbinn1985}). One may note here that transiting is 
a necessary condition for chaos, since orbits lying entirely within the sphere 
of influence (i.e. the so-called `pyramids' (\cite{merrvasil2011})) obey 
three quasi-integrals of motion which are deformations of the Keplerian 
integrals derived using perturbation theory (here, the perturbation 
consists of the triaxial galactic potential, which perturbs the otherwise 
Keplerian dynamics very close to the black hole, see \cite{merrvasil2011}). 
On reverse, as shown below, for transiting orbits one can determine 
the so-called {\it stretching number} (i.e. local Lyapunov number) 
yielding the local rate of deviation of two orbits with nearby initial 
conditions across one transit. The Lyapunov exponent, modeled as the sum 
of many such stretching numbers, turns then to be positive, i.e. the 
orbits are chaotic. We provide various types of evidence for the validity 
of this approximation, which allows to predict a (positive) mean value for 
the Lyapunov exponent as a function of the central mass parameter $m$. 
It is remarkable, in this respect, that during a transit the motion can 
be characterized as nearly Keplerian, while far from transits the motion 
is box-like and obeys three approximate integrals. Thus, both `piecewise' 
motions can be analyzed by nearly-integrable approximations. Nevertheless, 
their union yields chaos. 

In our basic modeling we use a simple galactic model with a harmonic 
core. This ensures a priori the presence of many box orbits before the 
insertion of the black hole. However, it is well known that realistic 
galactic models present power-law central density cusps $\rho(r)\propto 
r^{-\gamma}$ (\cite{feretal1994}, \cite{Lauer1995}, see review in 
\cite{binmer1998}, sect.4.3.1). Central cusps are characterized as 
`weak' if $\gamma<1$ or `strong' if $\gamma>1$, with the central force 
tending to zero or to infinity respectively. It is well known (e.g. 
\cite{mervalu1996}) that the central cusps are themselves an important 
source of chaos for centrophylic orbits. Thus, they significantly 
affect the value of $L$ even without the presence of a central black 
hole. We will show, however, by numerical tests, that the presence of 
the central cusp introduces a critical mass parameter scale $m_c$, 
depending on the value of $\gamma$, such that, for $m>m_c$ the 
chaotic scattering is dominated by the black hole, while for $m<m_c$ 
it is dominated by the central cusp. As shown in section 4, we then 
essentially recover an effective power law $L\propto m^p$ for $m>m_c$. 
Finally, we present numerical evidence that an effective power-law 
of the same form applies in rotating disc-barred galaxies with central 
black holes. In this case, the relevant mass parameter $m$ corresponds 
to the ratio of the mass of the black hole over the total mass of the 
bar. In summary, although our theoretical interpretation for the origin 
of the effective law $L\propto m^p$ is strictly valid in a quite simplified 
(and rather unrealistic) galactic model, our numerical evidence is that 
such a law appears generically in models of increasing complexity 
(and astrophysical interest). 

The structure of this paper is as follows: section 2 presents further 
numerical results about the empirical relation $L\propto m^p$, using 
a simple galactic-type potential to which we add the potential of the 
central mass. These results are used in order to probe numerically some 
aspects of subsequent modeling. Our theoretical modeling itself is 
presented in Section 3. Section 4 presents numerical results in 
models with central cusps and discusses the extent and limits of 
validity of previous results on the correlation between $L$ and 
$m$ in such models. Section 5 deals with the  $L\propto m^p$ relation 
in barred disc galaxy models, discussing both its applicability and 
origin, despite the non-existence, in such systems, of box-like orbits. 
Section 6 summarizes the main conclusions of the present study.

\section{NUMERICAL RESULTS}

\subsection{Hamiltonian model and numerical integrations}
At first we study the relation between $L$ and $m$ in a simple Hamiltonian 
model that captures the main features of motion near the center of an elliptical 
galaxy with non-singular central force field, to which we add a Keplerian term 
corresponding to the central mass. The Hamiltonian is:
\begin{eqnarray}\label{ham}
{H(x,y,z,p_x,p_y,p_z)}= {p_x^2\over 2}+{p_y^2\over
2}+{p_z^2\over 2} + V(x,y,z)
\end{eqnarray}
where $V=V(x,y,z)$ is the gravitational potential:
\begin{eqnarray}\label{vham}
{V(x,y,z)}= {1\over2} {\omega_1}^2 x^2+{1\over 2}{\omega_2}^2 y^2+{1\over
2}\omega_3^2z^2 + \varepsilon(xz+xy+y^2)^2-{m/ \sqrt{r^2+d^2}}~~.
\end{eqnarray}
The variables $(x,y,z)$ are cartesian position coordinates, $(p_x,p_y,p_z)$ 
their conjugate momenta, and $r=\sqrt{x^2+y^2+z^2}$. The softening parameter 
$d$ was added in the Keplerian potential in order to avoid large numerical 
errors when the orbits pass very close to the center. We selected a set of 
incommensurable frequencies $\omega_1=1$, $\omega_2=\sqrt{2}$, 
$\omega_3=\sqrt{3}$ so as to avoid dealing with resonant orbits satisfying some 
low-order commensurability. The anharmonicity parameter $\varepsilon$ was given 
values between 0.01 and 0.5. Various tests of the robustness of our results 
against $\varepsilon$ are reported below. We also note that the quartic 
potential term was choosen so as to represent a generic form without 
particular symmetries, while in galaxies with one or more planes 
of symmetry the potential presents a corresponding even symmetry. 

The gravitational potential (\ref{vham}) corresponds to a galaxy 
model with a harmonic core. The far more realistic case in which a central 
cusp is present is examined in detail in section 4 below. Here, the choice 
of the potential ensures a priori the existence of many regular box orbits, 
when $m=0$, which are transformed to chaotically scattered orbits when 
$m\neq 0$. In the model (\ref{vham}), the force grows linearly with 
distance from the center, while the force from the central mass falls 
as an inverse square law. The two forces become similar in magnitude at 
distances comparable to the radius (\cite{gerhbinn1985}):
\begin{equation}\label{rm}
{r_m}={m^{1/3}r_c}~~.
\end{equation}
The sphere $r=r_m$ is called sphere of influence of the central mass. 
The parameter $r_c$ is of order unity in units in which  the frequencies 
$\omega_x$, $\omega_y$, $\omega_z$ are of order unity (as in Eq.(\ref{ham})). 
Then, for the total mass of the galaxy one also has $M\sim r_c^3=O(1)$. The 
periods of orbits reaching apocentric positions $r_a>>r_m$ are of order 
$T\sim 2\pi$. 

The equations of motion resulting from the Hamiltonian (\ref{ham}) are:
\begin{eqnarray}\label{eqmot}
\dot{x}&=&\nonumber p_x \\
\dot{y}&=& \nonumber p_y \\
\dot{z}&=& \nonumber p_z \\
\dot{p_x}&=&\nonumber -\omega_1^2 x-2\varepsilon(xz+xy+y^2)(z+y)
+ {m x}/{(r^2+d^2)}^{3/2}\\
\dot{p_y}&=&\nonumber -\omega_2^2 y-2\varepsilon(xz+xy+y^2)(x+2y)
+ {m y}/{(r^2+d^2)}^{3/2}\\
\dot{p_z}&=& -\omega_3^2 z-2\varepsilon(xz+xy+y^2)x
+ {m z}/{(r^2+d^2)}^{3/2}~~~.
\end{eqnarray}
In Lyapunov exponent computations we also make use of the variational 
equations:
\begin{eqnarray}\label{eqvar}
\delta\dot{x}&=&\nonumber \delta p_x \\
\delta\dot{y}&=&\nonumber \delta p_y \\
\delta\dot{z}&=&\nonumber \delta p_z \\
\delta\dot{p_x}&=&-{{\vartheta^2 V}\over {\vartheta x}^2}\delta
x-{{\vartheta^2 V}\over {\vartheta x \vartheta y}}\delta y-
{{\vartheta^2 V}\over {\vartheta x\vartheta z}}\delta z\\
\delta\dot{p_y}&=&\nonumber-{{\vartheta^2 V}\over {\vartheta
y\vartheta x}}\delta x -{{\vartheta^2 V}\over { \vartheta
y}^2}\delta y-
{{\vartheta^2 V}\over {\vartheta y \vartheta z}}\delta z \\
\delta\dot{p_z}&=&\nonumber-{{\vartheta^2 V}\over {\vartheta
x\vartheta z}}\delta x- {{\vartheta^2 V}\over {\vartheta y \vartheta
z}}\delta y- {{\vartheta^2 V}\over {\vartheta z}^2}\delta z~~.
\end{eqnarray}
In numerical computations we solve together the equations (\ref{eqmot}) 
and (\ref{eqvar}). We use a 7-8th order Runge-Kutta method with 
fixed time step $\Delta t=10^{-4}$. For a choice of softening $d=10^{-3}$, 
this time step ensures that energy is conserved to within an error 
between $10^{-12}$ and $10^{-10}$ for all orbits.

The Lyapunov characteristic number (LCN) is defined through the 
relation:
\begin{equation}\label{LCN}
LCN=\lim_{t\rightarrow\infty}{1\over t}\ln\left(\xi(t)/\xi(0)\right)
\end{equation}
where $\xi(t)=\sqrt{\delta x(t)^{2}+\delta y(t)^{2}+\delta
z(t)^{2}+\delta p_x(t)^{2}+\delta p_y(t)^{2}+\delta p_z(t)^{2}}$.
In practical computations we make use of the finite time Lyapunov 
characteristic number:
\begin{equation}\label{chit}
\chi(t)={1\over t}\ln\left(\xi(t)/\xi(0)\right)~~.
\end{equation}

We made computations for ensembles of $n=200$ orbits for the mass parameter 
values $m=10^{-5}$, $3\times 10^{-5}$, $10^{-4}$, $3\times 10^{-4}$, $10^{-3}$, 
$3\times 10^{-3}$, $10^{-2}$, as well as the values of the anharmonicity 
parameter $\varepsilon = 0.01$, $0.1$ and $0.5$. The initial conditions of 
each ensemble are choosen as follows. For each initial condition, we choose 
first randomly a value of the energy in the range $0\leq E\leq 0.2$, with 
uniform distribution. Then, we produce an orbit of zero initial angular 
momentum, by setting $p_x=p_y=p_z=0$ and by solving for $r$ the equation 
$E=V(r,\theta,\phi)$, where $(r,\theta,\phi)$ are spherical coordinates 
corresponding to the cartesian point $(x,y,z)$, and $\cos\theta$, $\phi$ 
are choosen randomly with a uniform distribution in the intervals $[-1,1]$ 
and $[0,2\pi)$ respectively. The selected range of energies represents 
motions with apocentric distances of order $r\approx r_c$. These are  
all centrophilic orbits with a zero mean value of either component of 
the angular momentum. 

\subsection{Results}
\begin{figure}
\centering
\includegraphics[scale=0.65]{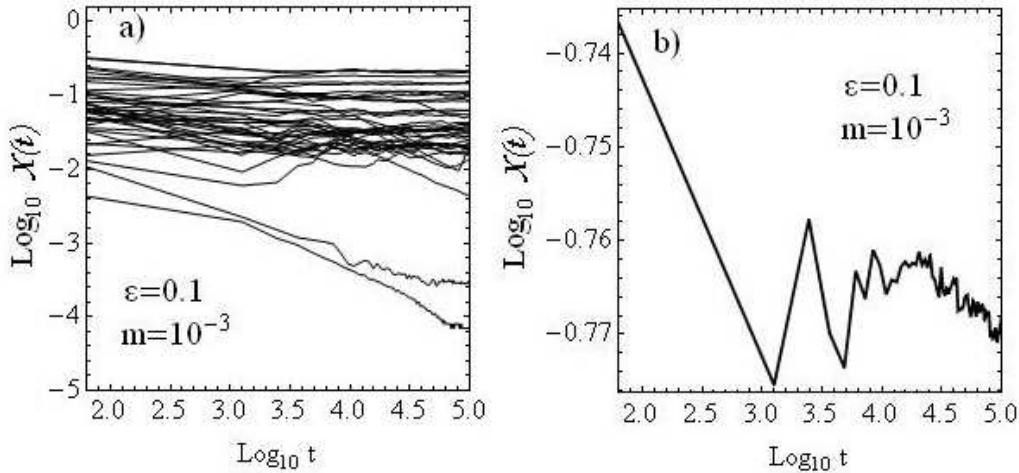}
\caption{ (a)Time evolution of finite Lypunov numbers for the ensemble 
of orbits in the case $m=10^{-3}$, $\varepsilon=0.1$. (b) The detailed form 
of the curve $\chi(t)$ for one chaotic orbit (initial conditions $x=0.064$, 
$y=-0.0625$, $z=-0.123$, $v_x=0.064$, $v_y=0.028$, $v_z=-0.06$. 
After the time $t=10^4$ the variations of $\chi(t)$ are less than 2\%.}
\label{liapt200}
\end{figure}

Figure \ref{liapt200} shows the time evolution of the finite-time 
Lyapunov exponent $\chi(t)$ for the whole ensemble of orbits in one 
of the numerical experiments where $m=10^{-3}$ and $\varepsilon=0.1$. 
The integration was up to the time $t=10^5$.
Almost all orbits in this ensemble are chaotic, as their finite-time 
Lyapunov exponents are stabilized to non zero values. Only a small
subset of orbits exhibit a value of $\chi(t)$ that keeps falling 
even at the time $t=10^5$, following the law $\chi(t)\sim 1/t$. 
This subset defines regular orbits. The detailed time behaviour of 
$\log\chi(t)$ for a typical chaotic orbit is shown in Fig.\ref{liapt200}b. 
A general remark is that for the entire set of parameter values used 
in our experiments, the large majority of orbits in our ensembles turn 
to be chaotic. The minimum percentage of chaotic orbits ($67\%$) is 
observed in the experiment with the minimum values of $m$ and $\varepsilon$, 
i.e. $m=10^{-5}$ and $\varepsilon=0.01$. The classification of orbits as 
ordered or chaotic is based on a `Fast Lyapunov Indicator' criterion 
(\cite{froetal1997}), namely on whether the length $\xi(t)$ of the deviation 
vector is smaller, or larger, respectively, than a threshold value set 
equal to $\xi_{th}(t)=100t$. 

\begin{figure}
\centering
\includegraphics[scale=0.80]{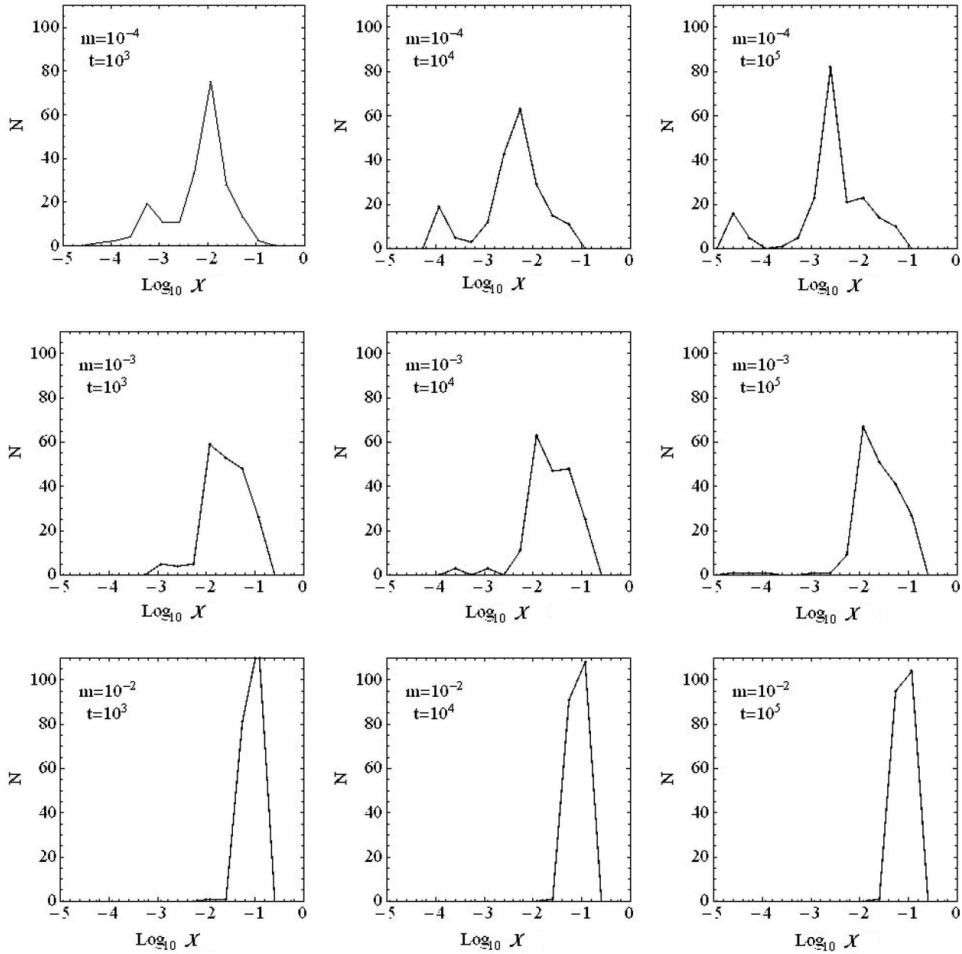}
\caption{The distributions  $P(\log \chi(t))$ for the particles in 
our ensembles. Every line of histograms corresponds to a different central 
mass parameter ($m=10^{-4}$, $m=10^{-3}$, $m=10^{-2}$) whereas every column 
to a different time snapshot ($t=10^{5}$, $t=10^{4}$, $t=10^{3}$). 
The histograms are shown for $\varepsilon$=0.1. }
\label{istogramliap}
\end{figure}
Figure \ref{istogramliap} shows the distribution of the quantity $\log\chi(t)$ 
for three different time snapshots ($t=10^3$, $t=10^4$, $t=10^5$) for 
the ensembles of orbits in the experiments with central mass values 
$m=10^{-4}$, $m=10^{-3}$ and $m=10^{-2}$. In all cases we can see that 
the distribution $\log{\chi(t)}$ exhibits a main peak corresponding to 
the chaotic orbits, which is displaced to higher values of $\log{\chi(t)}$ 
as the central mass $m$ increases. On the other hand, for the mass values  
$m=10^{-4}$ and  $m=10^{-3}$ there appears a secondary peak in the 
distribution of $\log\chi(t)$, that corresponds to the small subset of 
regular orbits. The secondary peak is displaced to the left, 
as the quantity $\log\chi(t)$ for regular orbits decreases in 
time as $-\log t$. For the largest central mass values ($m=10^{-2}$), 
however, we observe no secondary peak, i.e., all the orbits turn to 
be chaotic.

\begin{figure}
\centering
\includegraphics[scale=0.70]{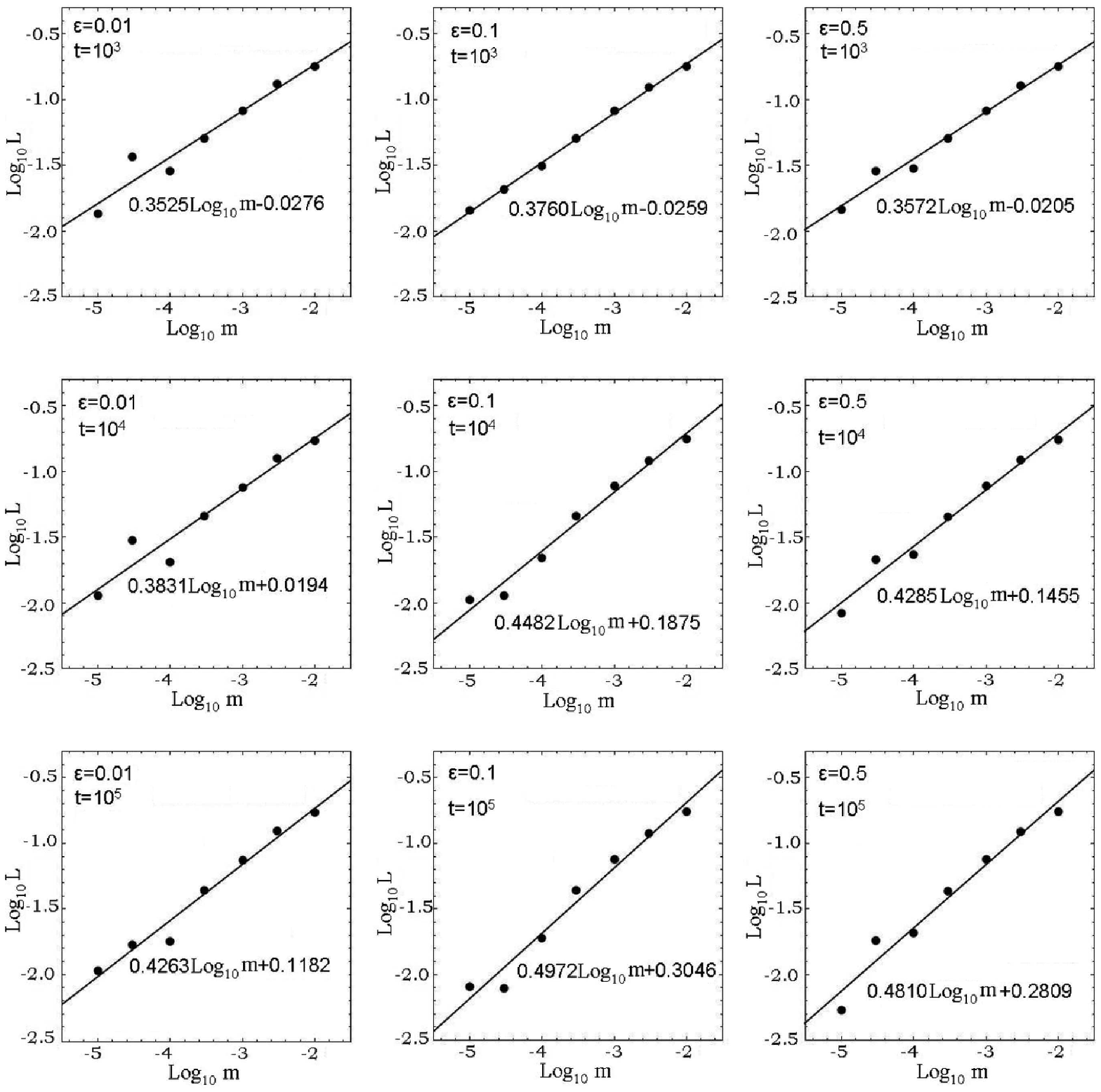}
\caption{The $L=\overline{\chi(t)}$ versus $m$ relation in logarithmic 
scale for all our experiments, with parameters as indicated in the 
panels. } 
\label{plloglm}
\end{figure}
Figure \ref{plloglm} shows the main result. From the histograms of 
Fig.\ref{istogramliap}, the mean value $L=\overline{\chi(t)}$ is 
extracted and plotted against $m$ at the snapshots $t=10^3$, 
$t=10^4$ and $t=10^5$ for all the numerical experiments. The 
straight lines in the same plots (in logarithmic scale) represent 
power-law fittings of the relation between $L$ and $m$. The best-fit 
exponents in different plots range in values between $p\simeq 0.35$ 
and $p\simeq 0.5$. We also note a tendency towards smaller values 
for smaller $t$. However, the bigger values are more representative 
of the true exponent, since they appear at times closer to the limit 
when $\chi(t)$ tends to its limiting value for chaotic orbits, i.e. 
the Lyapunov characteristic number (LCN). If 
$l=\lim_{t\rightarrow\infty}\chi(t)$ denotes the LCN limit, it is 
well known (see e.g. \cite{kalvog2005}) that the generic behavior 
of $\chi(t)$ is to fall like $t^{-1}$ up to a time $t_l\approx l^{-1}$. 
The time $t_l$ is called `Lyapunov time', and represents a saturation 
time beyond which the curve $\chi(t)$ starts stabilizing towards the 
limiting value $l$. The temporal change of $\chi(t)$ for times 
$t<t_l$ is reflected in the histograms of Fig.\ref{istogramliap}. 
We can observe that, along a fixed panel row (i.e. fixed $m$, 
integration time increasing from left to right), the right wing 
of the histogram is shifted in general to the left as $t$ increases. 
The shift is more conspicuous in the uppermost panel row (small $m$), 
in which the orbits have, in general, smaller values of the LCN, and 
hence, larger values of their saturation times $t_l$. On the 
other hand, in the lowermost panel row ($m=10^{-2}$, i.e., large), 
the saturation time is small ($t_l<10^3$ for nearly all orbits). 
Hence, the histogram $N(\chi)$ remains practically invariant beyond 
the time $t=10^3$, as shown in the three panels of the same row.

The difference in the saturation times between small and large $m$ has, 
as a consequence, that in each {\it column} of Fig.\ref{plloglm} (same 
parameters but increasing $t$), the value of $L$ for all $m$ presents 
some shift downwards as $t$ increases. The shift is important for 
small values of $m$, while it is nearly negligible for large values 
of $m$ (for which the orbits reach their saturation times $t_l$ already 
before $t=10^3$). Hence, the effective logarithmic slope $p$ increases 
as $t$ increases. Nevertheless, the tendency of $p$ to increase with 
time is only temporary, since $p$ is stabilized after all the orbits 
have reached their saturation times. This happens around $t=10^5$.  

\begin{figure} 
\centering
\includegraphics[scale=0.70]{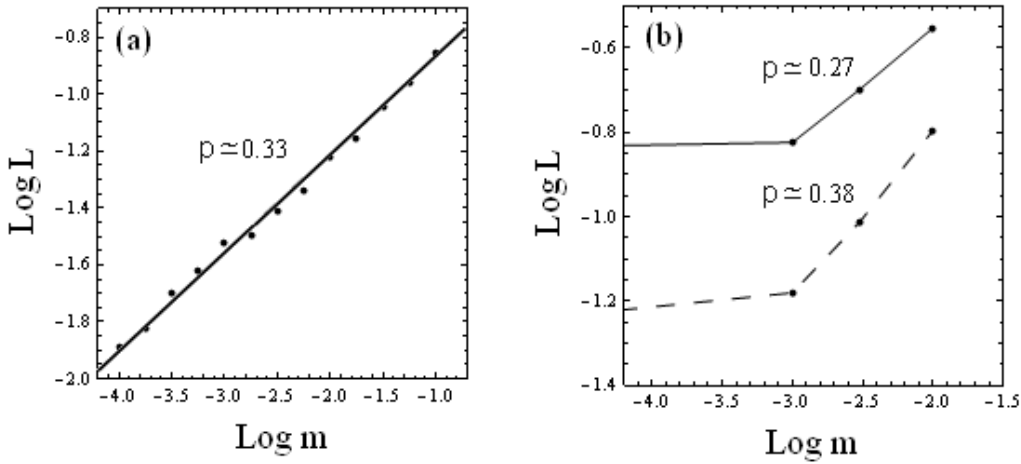}
\caption{The $L$ versus $m$ relation as derived from the data of 
(a) Kandrup and Sideris (2002) (lowest order approximation of 
a triaxial Dehnen potential), and (b)  Merritt and Valluri (1996) 
(potential corresponding to a galaxy with cusp). In (a), the 
compiled data confirm a power law relation $L\sim m^p$ over the 
whole range of values of $m$ considered. The solid line corresponds 
to a best fit yielding an exponent $p\simeq 0.33$. In the right 
panel (b) the solid and dashed lines show the dependence on $m$ 
of the logarithms of the mean first and second Lyapunov exponents 
respectively. Both lines extend to a non-zero value of $\log L$ 
for $m\rightarrow 0$. This represents the values corresponding 
to what is called `residual chaos' in section 4 below, i.e., chaos 
induced by the central cusp itself. The best-fit exponents 
were computed in this case only by the three rightmost data points.} 
\label{mervalkand}
\end{figure}
Another remark is that the dependence of the exponent $p$ on the 
anharmonicity parameter $\varepsilon$ appears to be weak. This fact 
confirms that the main source of the chaotic behavior of the orbits 
is the scattering by the central mass, while nonlinear effects due to 
the quartic terms in the potential are of small importance. This agrees 
with findings in (\cite{kandrsid2002}). It should be noted, in this 
respect, that a power-law relation $L\sim m^p$ can be extracted by 
an a posteriori analysis of data in independent published works  
(\cite{kandrsid2002}, \cite{mervalu1996}). In particular, \cite{kandrsid2002} 
computed finite-time Lyapunov Characteristic exponents in a potential 
representing the lowest expansion terms of a Dehnen potential with a 
superposed softened Keplerian term corresponding to the central mass. 
From their work (their figure 12), the relation between $L$ and $m$ can 
be compiled in a log-log scale. As shown in Fig.\ref{mervalkand}a, one 
obtains a power law fitting with $p\simeq 0.33$. Similarly, \cite{mervalu1996} 
computed the first and second (finite-time) Lyapunov characteristic 
numbers of chaotic orbits in potentials corresponding to a galaxy with 
cuspy density profile and a central core radius. Their results can be 
compared to ours in the limit where the core radius (their parameter 
$m_0$) is larger than the sphere of influence of the central mass. 
This is the case $m_0=10^{-1}$ in Table 1 of \cite{mervalu1996}, 
compiled in log-log scale in Fig.\ref{mervalkand}b. Apart from the 
value $m=0$ (corresponding to the horizontal line going to $-\infty$), 
the three available data points appear also to be aligned in straight 
lines indicating power laws both for the first and the second Lyapunov 
exponent. The best-fit exponents are $p\simeq 0.27$ and $p\simeq 0.38$ 
respectively, while these values are only indicative due to the scarcity 
of data and the unknown influence of the central cusp of the potential 
to the result. We note, finally, that the breaking of the power 
law and the appearance of a non-zero value of $L$ as $m\rightarrow 0$ in 
Fig.\ref{mervalkand}b is due to a phenomenon called below `residual chaos', 
i.e. chaos due to the cusp itself. This phenomenon is analyzed in 
section 4.

As an overall conclusion, a power-law $L\sim m^p$ appears quite 
commonly in numerical computations of the Lyapunov exponents of the 
stellar orbits chaotically scattered by a central mass in various 
galactic models. We now proceed in a theoretical modelling allowing 
to interpret the origin of this power law.

\section{Theoretical modelling}

\subsection{Transit and out-of-transit dynamics}
As mentioned in the introduction, a modelling of the process of chaotic 
scattering of the orbits by the central mass becomes feasible by considering 
two distinct regimes of the motion, i.e. i) transits through the sphere 
of influence and ii) out-of-transit oscillations. We begin by showing that 
within the out-of-transit regime the orbits obey three quasi-integrals 
of motion. Such integrals can be written as formal series, using, for 
example, the `third integral' approach (\cite{cont1960}). A formal series 
has the form $\Phi= \Phi_2+\Phi_3...$, where $\Phi_r$ are polynomial 
terms of degree $r$ in the canonical variables $(x,y,z,p_x,p_y,p_z)$. 
The term $\Phi_2$ is set equal to the harmonic energy in one of the 
three possible degrees of freedom, i.e. $(p_x^2 + \omega_x^2 x^2)/2$, 
$(p_y^2 + \omega_y^2 y^2)/2$, or $(p_z^2 + \omega_z^2 z^2)/2$. Terms 
of higher degree are computed recursively, by solving, order by order, 
the equation
\begin{equation}\label{inte}
{d\Phi\over dt}=[\Phi,H_2+H_4]=0
\end{equation}
where $[\cdot,\cdot]$ denotes the Poisson bracket operator, and $H_2$, 
$H_4$ are the quadratic and quartic terms of the Hamiltonian (\ref{ham}). 
This means that the formal integrals are possible to define for the 
Hamiltonian neglecting the influence of the central mass. Then, we 
test numerically how well they are preserved in the full model, 
in the out-of-transit regime. 

\begin{figure}
\centering
\includegraphics[scale=0.5]{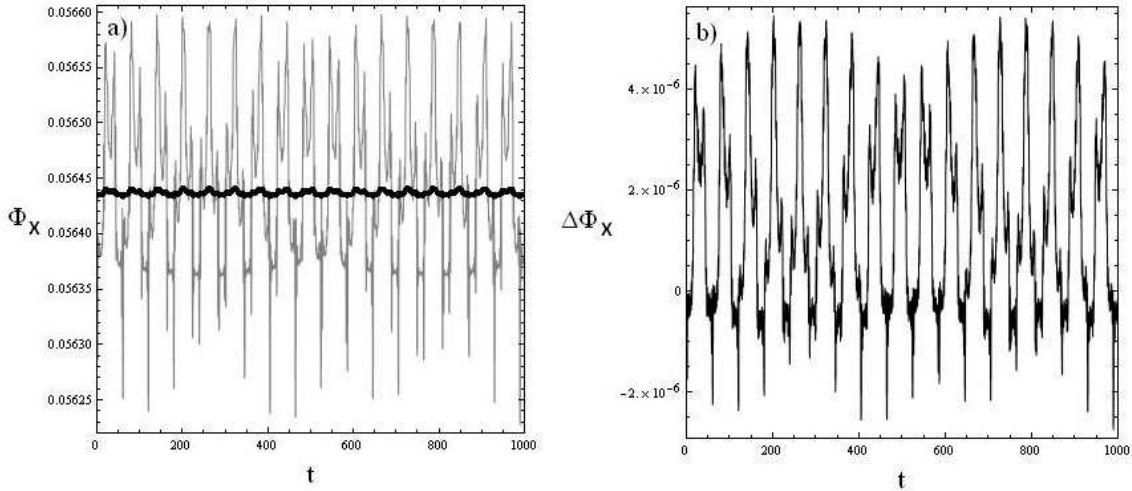}
\caption{(a) Variations of the value of the formal integral $\Phi_x$ 
truncated at order 4 (thin line) or 10 (thick line) for an example of box 
orbit with initial conditions $x=0.34$, $y=0.23$, $z=0.21$, $v_x=v_y=v_z=0$. 
(b) The variations of the order 10 truncated series in greater detail 
(notice the change of scale in the vertical axis).} 
\label{intevar}
\end{figure}
Equation (\ref{inte}) yields, at degree $r$, the homological equation
\begin{equation}\label{homo}
[{H_2,\Phi_r}]=[{\Phi_{r-2},H_4}]~~.
\end{equation}
We used a computer-algebraic program to solve, step by step, the homological 
equation, for all three formal integrals defined as the above way, up to 
the 10th degree in the variables $(x,y,z,p_x,p_y,p_z)$. As an example, 
for the formal integral $\Phi_x=(p_x^2 + \omega_x^2 x^2)/2+...$ up to 
4th degree we have
\begin{eqnarray}\label{}
{\Phi_x}&=&\nonumber {0.5x^2+0.5 p_x^2}
-\varepsilon\bigg(0.0025 p_x^2 p_y^2 - 0.007p_x p_y^3- 0.0034 p_x^2 p_y p_z
 \\&-&\nonumber 0.01 p_x p_y^2 p_z - 0.00125 p_x^2 p_z^2 
+ 0.0025 p_y^2 x^2+0.0034 p_y p_z x^2
 \\&+&\nonumber 0.00125 p_z^2 x^2 - 0.01 p_x p_y x y 
-0.007 p_y^2 x y-0.0052 p_x p_z x y
 \\&-&\nonumber 0.02 p_y p_z x y -0.0176 p_x p_y y^2 
+ 0.01 p_x p_z y^2- 0.0152x y^3
 \\&-&\nonumber 0.0086p_x p_y x z + 0.01 p_y^2 x z 
- 0.005 p_x p_z x z- 0.0017 p_x^2 y z
 \\&-&\nonumber0.04 p_xp_yyz + 0.0017 x^2yz- 0.03xy^2z 
- 0.00125p_x^2z^2
 \\&-& 0.00125 x^2z^2\bigg)+O(\varepsilon^2)~~.
\end{eqnarray}
Similar expressions are found for the formal integrals $\Phi_y$, $\Phi_z$. 
The degree of approximation of these expressions can be tested by 
probing how well the integral values are preserved along individual 
orbits integrated first in the Hamiltonian $H=H_2+H_4$ (i.e. without 
the central mass). For orbits in the energy range considered, we 
find that, up to $r=10$, all three integrals are preserved to within 
a time variation of about $10^{-(4+r/2-2)}$ at the truncation order $r$. 
Figure \ref{intevar} shows an example of this behavior. Panel (a) 
shows a comparison of the variations of the quantity $\Phi_x$, computed 
as above, at the truncation orders $r=4$ and $r=10$, for an example of 
box orbit with initial conditions as indicated in the panel. The maximum 
variation is about $\pm 2\times 10^{-4}$ at the truncation order $r=4$, 
but it reduces to about $\pm 3\times 10^{-6}$ at the truncation order 
$r=10$ (magnified in panel (b)). In fact, the estimates of Nekhoroshev 
theory (see, e.g., \cite{eftetal2004}) yield that the variations continue 
to decrease up to an optimal truncation order $r_{opt}\sim 1/(\varepsilon E)$, 
in which they become exponentially small, i.e., of order 
$O(\exp(-1/(\varepsilon E))$. However, even low order truncations 
are sufficient for estimating the values of the integrals $\Phi_x$, 
$\Phi_y$ and $\Phi_z$ in practice. 

\begin{figure}
\centering
\includegraphics[scale=0.6]{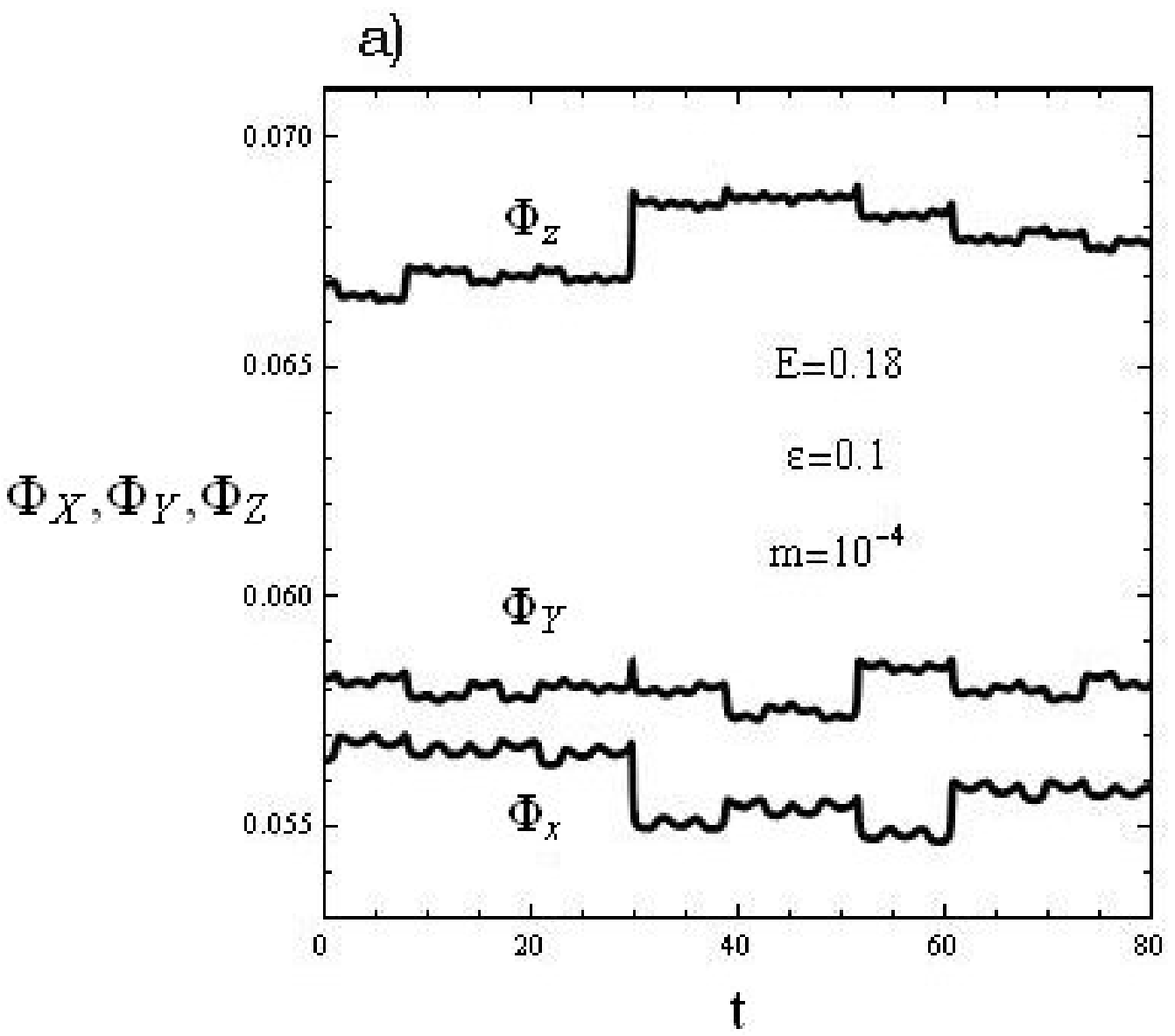}
\includegraphics[scale=0.39]{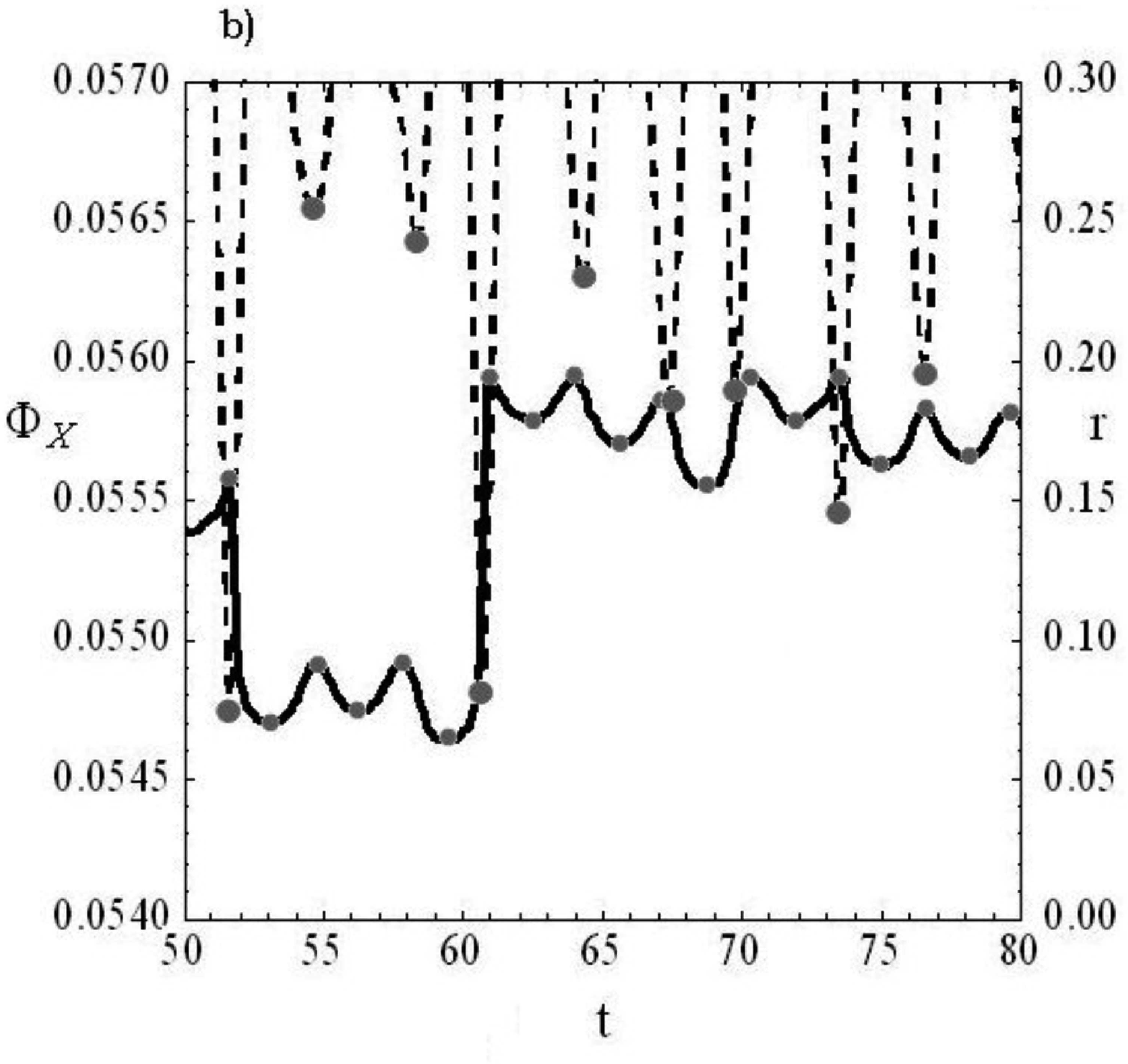}
\caption{(a) Time evolution of the quantities  $\Phi_x, \Phi_y, \Phi_z$ 
for one orbit with energy $E=0.18$ in the experiment with $m=10^{-4}$, 
$\varepsilon=0.1$ (initial conditions as in Fig.\ref{intevar}). 
(b) The time evolution of $\Phi_x$ in detail 
(solid curve, corresponding to the scale of the left vertical 
axis), along with the time evolution of the orbit's distance from the 
center (dashed curve, corresponding to the scale of the right vertical 
axis). Gray points on the latter curve correspond to local minima, 
while gray points on the $\Phi_x$ curve correspond to local extrema.  
Note that a local maximum of the curve $\Phi_x(t)$ coincides always 
with a local minimum of the curve $r(t)$.} 
\label{phijump}
\end{figure}
Restoring, now, the term $-m/(r^2+d^2)^{1/2}$ in the potential, we 
compute the variation of all three quantities $\Phi_x, \Phi_y, \Phi_z$ 
during both transits and the out-of-transit regime. Figure \ref{phijump}a 
shows the example of an orbit with initial conditions as in Fig.\ref{intevar} 
(energy $E=0.18$) in the case $m=10^{-4}$, $\varepsilon=0.1$. 
All three integrals are seen to exhibit abrupt jumps 
that coincide in time (e.g. at the times $t=30$, $t=52$ or $t=61$ in 
Fig.\ref{phijump}a). A closer focus to the jump at $t=52$ is shown in 
(Fig.\ref{phijump}b), superposing the variations of $\Phi_x$ in time with 
the variations of the distance of the orbit from the center. Clearly, 
the most important jumps take place during transits through the sphere 
of influence of the central mass (which is about $2\times 10^{-4/3}
\simeq 0.1$ in this case, see below, end of subsection 3.1). 
A similar behavior is found for the quasi-integrals $\Phi_y$, $\Phi_z$. 
On the other hand, the values of all three integrals are preserved to 
within two significant figures in the out-of-transit regime (without 
the central mass, the precision increases to about four significant 
figures at the truncation order $r=4$, see above). 

\begin{figure}
\centering
\includegraphics[scale=0.8]{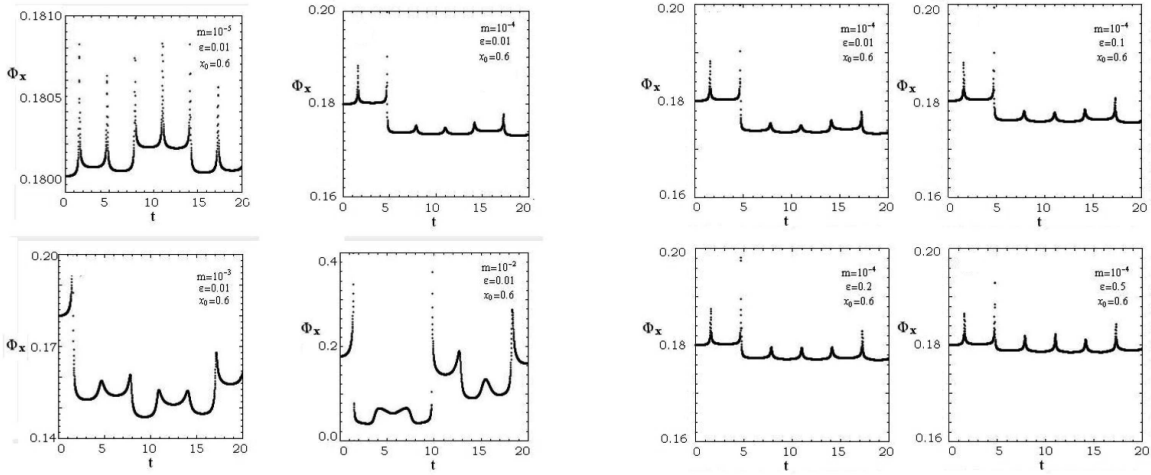}
\caption{Variations of $\Phi_x(t)$ for an orbit with energy $E=0.18$ and 
four different values of $m$ (left), or $\epsilon$ (right). The values 
of $(\mu,\varepsilon)$ are indicated in the figure. Note the change of 
scale in the ordinate of the four left panels, indicating that the 
variations of the quasi-integral depend strongly on the value of the 
mass parameter (while they are nearly insensitive to the non-linearity 
parameter $\varepsilon$ as long as the latter is not very close to 
unity).}
\label{phivarpar}
\end{figure}
Figure \ref{phivarpar}, now, compares the variations of $\Phi_x(t)$ for an orbit 
with the same initial conditions but different values of $m$ and $\varepsilon$. 
The overall size of the variations appears rather insensitive on $\varepsilon$, 
while the size of all jumps (i.e. the difference between the value of $\Phi_x$ 
at the local maximum and minimum along a jump) clearly increases as $m$ 
increases. One can observe also that the `landing' value of $\Phi_x(t)$ 
at the end of one jump appears to be more and more unpredictable as $m$ 
increases. Essentially, this uncertainty in the value of $\Phi_x(t)$ 
after each jump is a measure of the chaoticity of the orbit.

\begin{figure}
\centering
\includegraphics[scale=0.8]{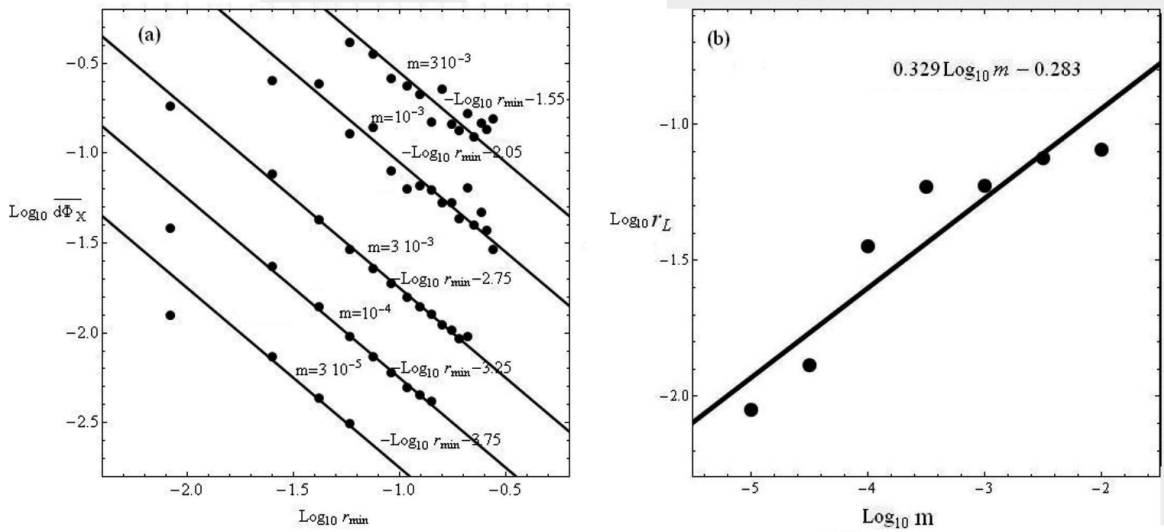}
\caption{ (a) The points represent the logarithm of the mean absolute 
value of the $\Phi_x$ integral's jumps $\log\overline{d\Phi_x}$ for different 
values of the central mass parameter, versus the logarithm of the mean minimum 
distance $r_{min}$ from the central mass over all transits for chaotic orbits 
with initial conditions as in Fig.\ref{phijump}.  The fitting straight lines 
have equation as indicated in the figure. (b) The mean radius $r_L$ within 
which the angular momentum is approximately conserved (see text) versus $m$.} 
\label{plkep}
\end{figure}
Quantifying the behavior of the jumps after many transits allows to 
see that the scattering of the orbits can be modelled essentially by  
Keplerian hyperbolic dynamics. Figure \ref{plkep}a shows the mean value of the 
jumps $d\Phi_x$ (measured as the difference between the local maximum and 
minimum values of $\Phi_x$ at each jump) plotted against the mean value of 
the radii $r_{min}$, where $r_{min}$ (computed from the data of 
Fig.\ref{phijump}b) means the radial distance from the central mass at 
the point of closest approach during one jump. Both means are taken over 
the ensemble of all jumps during an integration of an orbit up to a time 
$t=10^5$. The computation is repeated for different central mass values 
$m$, keeping the orbit's initial conditions fixed (same as in Fig.\ref{phijump}). 
During this time, the orbit exhibits some thousands of transits, 
thus the evaluation of mean values for both $d\Phi$ and $r_{min}$ has 
small statistical error. Figure \ref{plkep}a shows the mean value of $d\Phi_x$ 
as a function of the mean value of $r_{min}$, superposing the plots in log-log 
scale for the masses $m=3\times 10^{-5}$, $10^{-4}$, $3\times 10^{-4}$, 
$10^{-3}$, and $3\times 10^{-3}$. The straight fitting lines have inclination 
-1, whereas we note that the vertical distance between two successive 
lines is about equal to $\log_{10}3\approx 0.5$. Thus, the mean jump 
$\overline{d\Phi_x}$ is consistent with a (absolute) Keplerian potential 
law: 
\begin{equation}\label{dphi}
\overline{d\Phi_x}\approx {m\over \overline{r_{min}}}~~.
\end{equation}
The above result concerns the variations of the quasi-integrals valid 
in the out-of-transit regime. On the other hand, inside the sphere of 
influence the force field is approximately central, thus another 
approximate integral, of the angular momentum $\mathbf{L}$, is valid. 
During transits, we follow the time variations of $\mathbf{L}$ and 
determine, at each transit separately, the maximum radius $r_L$ up to 
which the variations $\Delta\mathbf{L}$ are smaller in size than a fixed 
percentage of $\mathbf{L}$ (taken as $10\%$ difference measured from 
the value of $\mathbf{L}$ at the point of closest approach to the 
central mass). Figure \ref{plkep}b shows the so-obtained mean value 
of $r_L$ as function of the central mass parameter $m$ in log-log scale 
for the same orbits as in Fig.\ref{plkep}a. Albeit with considerable 
scatter, the data allow to determine a best-fit power law 
$r_L\approx 2.2m^{0.329}$. This is close to the power law for the sphere 
of influence (Eq.(\ref{rm}), thus allowing to identify $r_L$ as a measure 
of $r_m$ yielding the estimate $r_c\approx 2$. We note that the 
threshold of 10\% variation of the angular momentum is rather arbitrary. 
However, it allows to obtain a meaningful result for box orbits which, 
far from the BH sphere of influence, undergo angular momentum variations 
of order 100\% (with more stringent thresholds we identify significantly 
fewer transitions through the BH sphere of influence).

\begin{figure}
\centering
\includegraphics[scale=0.50]{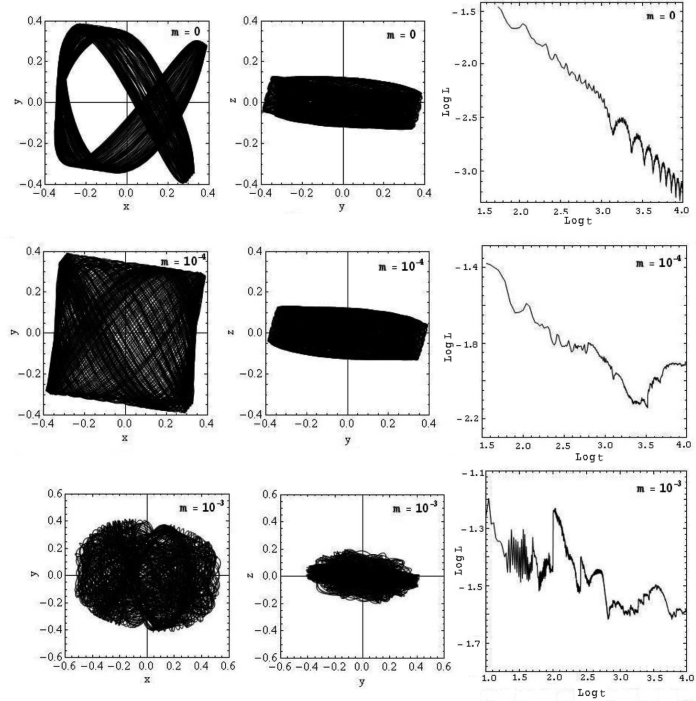}
\caption{ Upper row: projection of a thin tube orbit around a `boxlet' 
3:2 periodic orbit in the planes $(x,y)$ (left panel), and $(y,z)$ 
(middle panel), along with the evolution of its finite-time Lyapunov 
exponent $\chi(t)$ (right panel). The central mass parameter is set 
equal to $m=0$, and the nonlinearity parameter $\varepsilon=0.9$. 
The orbit's initial conditions are: $x=0.1894$, $y=-0.04796$, $z=0.0014$, 
$v_x=-0.3578$, $v_y=-0.4682$, $v_z=-0.097$. The orbital energy is $E=0.193$. 
Middle and lower rows: same as in the upper row, but with a central mass 
$m=10^{-4}$ and $m=10^{-3}$ respectively.}
\label{boxlet}
\end{figure}
In conclusion, their scatter notwithstanding, the previous plots 
suggest a qualitative picture in which the dynamics of chaotically scattered 
centrophylic orbits can be modelled as a sequence of i) transits 
through the sphere of influence, in which the orbits follow approximately 
a Keplerian hyperbolic dynamics, followed by ii) box-like wanderings 
within the rest of the available space in the interior of the equipotential 
ellipsoid corresponding to a fixed value of the orbital energy.  
This picture is quite generic when the three frequencies $\omega_1$, 
$\omega_2$, $\omega_3$ are far from low-order commensurabilities. 
The generation of such commensurabilities necessitates a separate 
treatment, since, then, the orbital sample contains also many resonant 
boxlets (\cite{miralda1989}). In our model, boxlets corresponding to 
low-order commensurabilities are generated by the quartic potential 
terms in Eq.(\ref{vham}) for large (${\cal O}(1)$) values of $\varepsilon$. 
An example is given in Fig.\ref{boxlet}, for $\epsilon=0.9$. When 
$m=0$ (top row) the orbit with initial conditions as indicated in the 
figure's caption is a three-dimensional thin-tube orbit around a resonant 
$3:2$ orbit which exists in the plane $(x,y)$. The periodic orbit is stable, 
and, hence, centrophobic (\cite{mervalu1999}). In Fig.\ref{boxlet}, top row, 
the tube orbit around the boxlet also avoids the center. Hence, it is a 
regular orbit, as confirmed by computing its Lyapunov characteristic 
exponent (right panel), which goes to zero. Now, the orbit's closest 
approach to the center is at a distance $r\sim 2\times 10^{-2}$. Thus, 
by adding a central mass with parameter $m=10^{-4}$ (middle row), the orbit 
now crosses the central mass sphere of influence (of radius $\sim 10^{-4/3}$). 
As a result, we observe that the orbit is significantly deformed, and looses 
its resonant character, while the Lyapunov exponent stabilizes to a positive 
value $\sim 10^{-2}$, i.e., the orbit becomes weakly chaotic. For still 
larger $m$ ($m=10^{-3}$), the orbit exhibits the usual behavior of a chaotic 
centrophylic orbit, with a Lyapunov exponent $\sim 2.5\times 10^{-2}$. 
 
We now model the chaotic scattering process of centrophylic orbits in order 
to derive theoretical estimates for the orbits' Lyapunov exponents.

\subsection{Theoretical Estimates on Lyapunov Exponents}                                  
Consider an orbit integrated along with the variational vector 
$\mathbf{\xi}(t)$ of a nearby orbit up to the time $t$. Let $\xi_i$ be 
the modulus of the deviation vector of the orbit at the time 
$t_i= i\Delta t$, where $\Delta t= t/n$ is the timestep corresponding 
to a splitting of the integration in $n$ steps. The {\it stretching 
number} \cite{vogcon1994} at the i-th step is defined as
\begin{equation}\label{alpha}
a_i={1\over\Delta t}\ln{\xi_i\over \xi_{i-1}}~~.
\end{equation}
The finite time Lyapunov characteristic number is equal to the mean 
stretching number along the orbit, since
\begin{equation}\label{chialpha}
\chi(t)={1\over t}\ln{\xi(t)\over \xi_0}
={1\over{n\Delta t}}\sum_{i=1}^{n}\ln{\xi_i\over{\xi_{i-1}}}
={1\over n}\sum_{i=1}^{n}a_i~~.
\end{equation}

\begin{figure}
\centering
\includegraphics[scale=0.7]{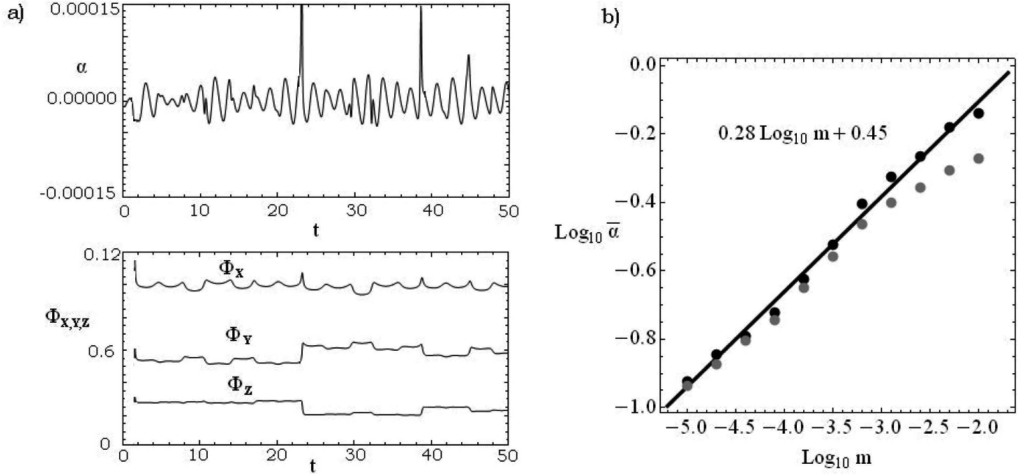}
\caption{(a) Comparison of the time evolution of the stretching number 
$a(t)$ (Eq.(\ref{alpha})) with the variations of the quasi-integrals 
$\Phi_x$, $\Phi_y$, $\Phi_z$. (b) The mean stretching number $\overline{a}$ 
after thousands of transits versus $m$ for the orbit with same initial 
conditions as in Fig.\ref{phijump}. A power-law fit yields $\overline{a}
\propto m^{0.28}$. The theoretical estimate $\overline{a}\propto m^{1/3}$ 
is found in the appendix.} 
\label{plstr}
\end{figure}
Figure \ref{plstr}a shows the time evolution of the stretching number 
$a(t)$ as a function of time, along with the variations of the integral 
$\Phi_x(t)$ for the orbit with initial conditions as in Fig.\ref{phivarpar}, 
for $m=10^{-3}$, $\varepsilon=0.1$. Far from transits, the function $a(t)$ 
shows an oscillatory behavior around zero. This behavior is characteristic 
of a nearly-harmonic oscillation, while the quartic potential term implies 
an overall linear growth of the deviation vector in the out-of-transit 
regime, which is of order $O(\varepsilon) t$. On the other hand, in 
Fig.\ref{plstr}a the curve $a(t)$ clearly exhibits positive peaks at  
all transits. 

Consider the set ${\cal S}=\{i_1,i_2,...,i_{n_T}\}$ such that the orbit 
is in transit at the time $t_i$ with $i\in S$, $n_T$ denoting the total 
number of timesteps during which the orbit is in the transit phase. 
Let $\overline{{\cal S}}$ be the complement of ${\cal S}$ with respect 
to the set $\{1,2,...,n\}$. One has the estimate:
$$
{1\over{n\Delta t}}\sum_{i_j\in\overline{{\cal S}},j=1}^{n-n_T}
\ln{\xi_{i_j}\over{\xi_{i_j-1}}}
\approx
O\left({\ln(\varepsilon n\Delta t)\over n\Delta t}\right)
$$
implying that the contribution of the `out-of-transit' stretching numbers 
to the final value of $\chi(t)$ goes to zero as $\sim\ln t/t$. Thus, one 
has the approximation 
$$
\chi(t)\approx
{1\over{n\Delta t}}\sum_{i_j\in{\cal S},j=1}^{n_T}a_{i_j}~~.
$$
Setting $\overline{a}=(1/n_T)\sum_{i_j\in{\cal S},j=1}^{n_T}a_{i_j}$ 
we get $\chi(t)\approx (n_T/n\Delta t)\overline{a}$. The quantity 
$$
N_{vis} = {n_T\over n\Delta t}
$$
hereafter called `rate or visits', represents the number of transits per 
unit period of an orbit whithin the sphere of influence. We then have
\begin{equation}\label{chialpha2}
\chi(t)\approx N_{vis}\overline{a}~~.
\end{equation}
Estimates on the mean value of $\chi(t)$ for all transiting orbits at fixed 
central mass parameter $m$ will then follow by estimating separately the quantities 
$\overline{a}$ and $N_{vis}$. 

Assuming, as evidenced above, that the transits are governed by nearly-Keplerian 
hyperbolic dynamics, in the Appendix it is shown that for orbits of given energy 
$E$, one has the theoretical estimate
\begin{equation}\label{meana}
\overline{a}\propto {m^{1/3}\over E^{1/2}}~~.
\end{equation}
Figure \ref{plstr}b shows the mean $a$ computed numerically for an orbit of 
fixed energy with the same initial conditions as in Fig.\ref{phijump}, 
integrated under various values of $m$. Numerically we find the exponent 
0.28, which is in fair agreement with the theoretical exponent $1/3$ of 
Eq.(\ref{meana}). The predicted dependence of $\overline{a}$ on the energy, 
probed numerically below, is also verified.  

\begin{figure}
\centering
\includegraphics[scale=0.5]{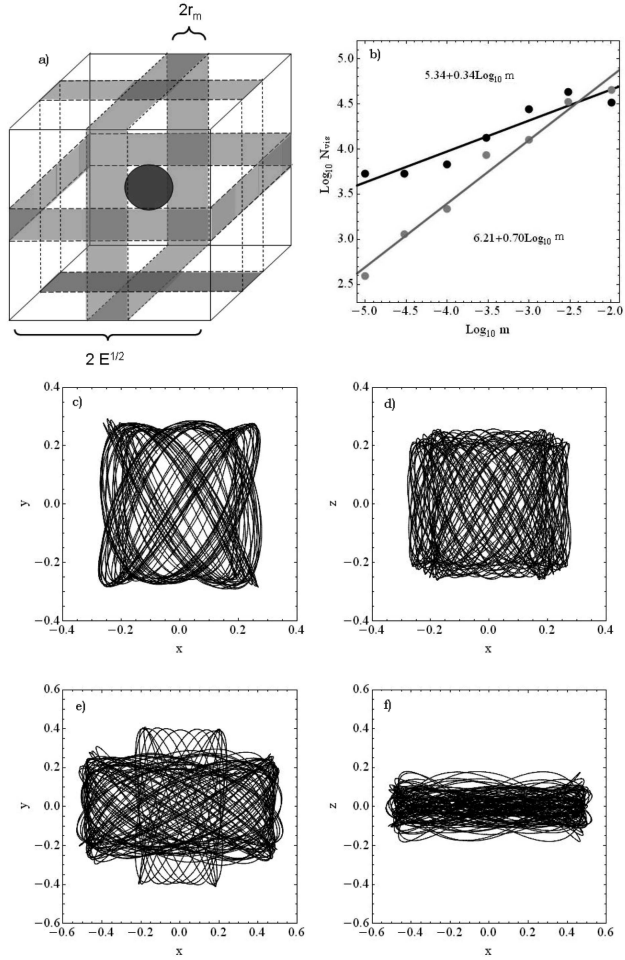}
\caption{(a) Schematic representation of the configuration space of motion 
of  ``3D-orbits'' and  ``planar'' orbits. The cube represents a box 
volume covered by a ``3D-orbit''. The central sphere represents the central 
mass sphere of influence.  The three thin parallelepipeds (gray) correspond 
to one parallelepiped side being equal to $2r_m$. Planar orbits are orbits 
lying inside one of the three parallelepipeds. (b) The number of visits 
(up to time $t=10^5$) versus $m$ for two orbits characterized as `3D' 
(black points, initial conditions $x=0.227$, $y=0.155$, $z=0.139$, 
$v_x=v_y=v_z=0$), and `planar' (gray points, initial conditions $x=0.227$, 
$y=0.155$, $z=0.0$,$v_x=v_y=v_z=0$), for $\varepsilon=0.1$. 
The straight lines represent power-law fittings yielding the best-fit 
exponents 0.7 and 0.34 respectively. (c-d) Projection of the `3D' orbit 
in the planes $(x,y)$ and $(x,z)$. (e-f) Same for the `planar' orbit. 
Note in (e) the change in morphology induced by a big jump in the values 
of the quasi-integrals $\Phi_x$, $\Phi_y$. Such jumps are stochastic in 
nature, and they may occasionally convert a 3D orbit to planar and vice
versa. } 
\label{schematic}
\end{figure}
We now focus on estimating $N_{vis}$. The frequency whereby an orbit visits 
the sphere $r_m$ depends on the geometry of the orbit in configuration space. 
We distinguish two cases, explained with the help of Fig.\ref{schematic}:

i) {\it 3D-orbit}: as long as the three quantities $\Phi_x$, $\Phi_y$, 
$\Phi_z$ obtain comparable values, an orbit fills nearly uniformly the 
available configuration space, which has the form of a deformed 3D 
box.

ii) {\it planar orbit}: at least one of the three quantities $\Phi_x$, 
$\Phi_y$, $\Phi_z$ obtains a value smaller than a given threshold 
(given by Eq.(\ref{orbplane}) below). Geometrically, the amplitude 
of oscillations in at least one of the three axes in the out-of-transit 
regime should be smaller than the radius $r_m$ of the sphere of influence  
(Fig.\ref{schematic}a, schematic). Quantitatively  
\begin{equation}\label{orbplane}
(2\Phi_k/\omega_k^2)^{1/2}<r_m
\end{equation}
where $k$ stands for $x$, $y$, or $z$. 
 
Note that `linear' orbits, i.e., tubes around the stable axial orbits 
also exist, but their importance is rather limited because they are 
considerably fewer than the planar or 3D orbits.

The dependence of the rate of visits to the central masses' sphere of 
influence on the geometry of orbits can be modelled in the following way: 
for 3D-orbits, considering all possible straight line segments connecting 
two different points on the surface of the box delimiting the orbit 
(see Fig.\ref{schematic}a), $N_{vis}$ can be approximated as proportional 
to the percentage or line segments passing through the sphere of influence. 
Then, $N_{vis}\propto S_{r_m}/S_{tot}$, where $S_{r_m}$ and $S_{tot}$ are 
the surface of the sphere of influence and of the box respectively. 
The linear dimension of the box is of order $l\sim E^{1/2}$, where $E$ 
is the energy of the orbit. Thus, $N_{vis}\propto r_m^2/E$, or (taking 
into account Eq.(\ref{rm}))
\begin{equation}\label{nvis3d}
N_{vis}\propto{m^{2/3}\over E}~~.
\end{equation}
For planar orbits, one has, instead, the estimate 
$N'_{vis}\propto r_m/l$, or
\begin{equation}\label{nvis2d}
N'_{vis}\propto{r_m\over E^{1/2}}={m^{1/3}\over E^{1/2}}~~.
\end{equation}
These estimates are confirmed numerically. Fig.\ref{schematic}b shows 
a computation of the rate of visits $N_{vis}$, $N_{vis}'$ for two orbits 
with constant energy $E=0.18$, but for different mass parameters $m$. 
The number of visits within a total integration time $t=10^5$ are counted, 
and criterion (\ref{orbplane}) is used in order to distinguish either orbit 
as `planar' or 3D (the corresponding rate of visits is found by 
dividing the number of visits by $t=10^5$). The difference in the shape 
of the orbits is evident (Figs.\ref{schematic}c-f). Returning to 
Fig.\ref{schematic}b, the best-fit exponents of the relations 
$N_{vis}$ and $N_{vis}'$ to $m$ are 0.7 and 0.34 respectively, 
in fair agreement with Eqs.(\ref{nvis3d}) and (\ref{nvis2d}) for 
constant $E$. 

\begin{figure}
\centering
\includegraphics[scale=0.7]{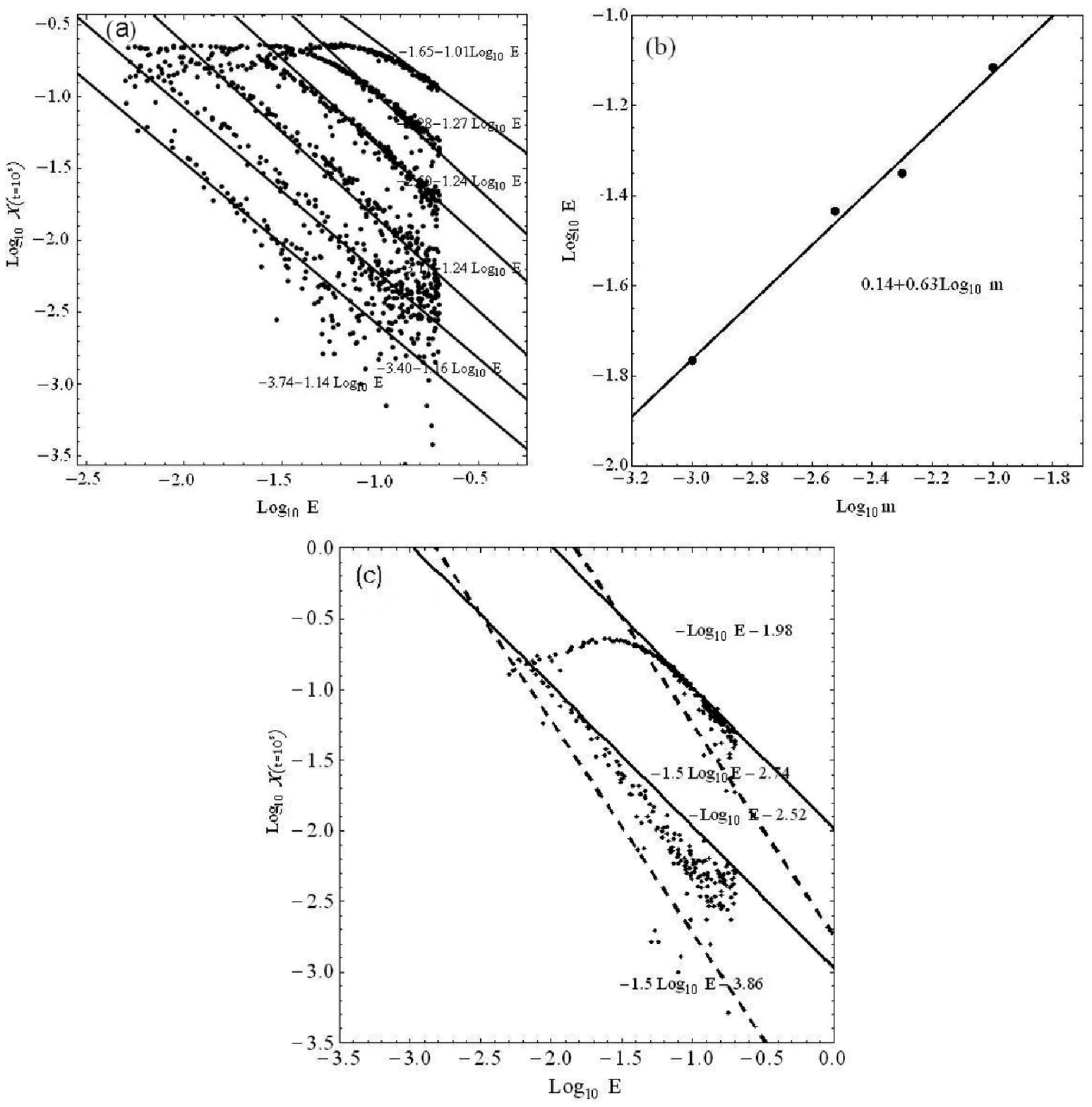}
\caption{(a) The logarithms of finite-time Lyapunov characteristic numbers 
$\log\chi$ of all orbits in the choosen ensembles versus the logarithm of 
the orbital energy $\log E$ for the experiments with mass parameters (from 
top to bottom) $m=10^{-2}$, $m=3\times10^{-3}$,  $m=10^{-3}$,  $m=3\times10^{-4}$,  
$m=10^{-4}$ and $m=3\times10^{-5}$, at the end of the integration ($t=10^5$). 
The straight lines represent asymptotic power-law fittings for the right 
parts of the plots separately for each mass parameter.  
(b) The energy $E_{max}$ where $\log\chi$ in (a) exhibits a global maximum 
versus $m$. The power-law fiting is $E_{max}=2.2{m^{2/3}}$. (c) The plot 
$\log{\chi}$ versus = $\log E$ in greater detail for the masses $m=3\times 
10^{-3}$, and $m=3\times 10^{-4}$. In each case, numerical values are 
distributed between two lines with inclination -1 and -1.5, as predicted 
from Eqs.(\ref{lyap2d}) and (\ref{lyap3d}) for the planar and the 3D 
orbits respectively.}
\label{logElogllogm}
\end{figure}
Taking into account eqs.(\ref{nvis2d}),(\ref{nvis3d}) and (\ref{meana}) we 
find for the Lyapunov number of 3D-orbits of energy $E$ the estimate:
\begin{equation}\label{lyap3d}
\chi \approx { m^{2/3}\over E} \times {m^{1/3}\over E^{1/2}}
\approx {m\over E^{3/2}}
\end{equation}
whereas, for planar orbits
\begin{equation}\label{lyap2d}
\chi'\approx {m^{1/3}\over E^{1/2}}\times {m^{1/3}\over E^{1/2}}
\approx {m^{2/3}\over E}~~.
\end{equation}
Figure \ref{logElogllogm}a shows the values of $\log\chi$ against $\log E$ 
for all the orbits in our considered ensembles for six different values of the 
mass parameter $m$ as indicated in the caption. The various ensembles are clearly 
distinguished by the concentration of points in the scatter plot, the uppermost 
concentration corresponding to the ensemble in the experiment with the highest 
mass parameter ($m=10^{-2}$). Each ensemble can be roughly described as 
consisting of a `rising' and a `falling' part of the value of $\log\chi$ 
vs. $\log E$. The two parts meet at a point of maximum of $\log\chi$. 
The position of the maximum moves to the right with respect to $\log E$ 
as $m$ increases. However, the level value of $\log\chi$ at the maximum 
remains remarkably constant, i.e. nearly independent of $m$. 

The straight lines show power law fittings of $\chi$ with $E$ for the falling 
part. Despite the large scatter of the data points, we find indicative 
logarithmic slopes lying in the range between -1 and -1.5 for all ensembles 
considered. This behavior will be explained below. A power-law roughly holds 
also in the rising part. The point of maximum corresponds to about the point 
where the associated best-fit power laws intersect. The intersection point 
defines an energy $E_c$. Computing and plotting $E_c$ against $m$ yields 
approximately a power-law $E_c\propto m^{0.63}$ (Fig.\ref{logElogllogm}b). 

These features can be understood by the following considerations: first,  
one can note that the left part represents regular or sticky chaotic orbits 
which have the morphology of {\it pyramids} (\cite{merrvasil2011}), i.e. 
they lie nearly entirely within the sphere of influence of the central mass. 
Such orbits can be described by perturbations to the Keplerian dynamics 
of the central mass. The limiting energy value $E_l$ up to which an orbit 
lies entirely within the radius $r=r_m$ can be estimated by requiring that 
the sphere $r=r_m$ constitutes the surface of zero-velocity. The estimate   
\begin{equation}\label{qwer}
E_{l} \approx {1\over 2}\Omega^2{r_{m}^2}-{m\over r_{m}}
\approx{1\over 2}\Omega^2r_c^2{m^{2/3}}-{m^{2/3}\over r_c}
=\left({r_c^2\Omega^2\over 2}-{1\over r_c}\right)m^{2/3}
\end{equation}
holds, where $\Omega$ is a geometric-mean estimate of the harmonic frequencies, 
bounded by the highest of the three frequencies $\omega_x,\omega_y,\omega_z$. 
The dependence of $E_l$ on $m$ in Eq.(\ref{qwer}) has the exponent $0.66$, 
close to the exponent in the numerical fitting of $E_c$ versus $m$ 
(Fig.\ref{logElogllogm}b). This suggests that $E_l\simeq E_c$ (the 
near equality holds also checking numerical vs. theoretical coefficients).
Use of the estimate (\ref{qwer}) is made in the next subsection.

On the other hand, most chaotic orbits lie beyond the energy $E_{max}$. 
In fact, isolating the plots $\log\chi$ vs. $\log E$ (Fig.\ref{logElogllogm}c) 
for two values of the mass parameter $m$ allows to see that the whole ensemble 
of orbits in the right wing are delimited between two limiting lines with 
inclinations -1 and -1.5 respectively, i.e. as predicted from the estimates 
of Eqs. (\ref{lyap2d}) and (\ref{lyap3d}) for the planar and the 3D orbits 
respectively. The coexistence of `planar' and `3D' orbits explains 
in this way the scatter in the data points of Fig.\ref{logElogllogm}a.

\subsection{Final theoretical estimates: the power law $L\propto m^p$}
\begin{figure}
\centering
\includegraphics[scale=0.7]{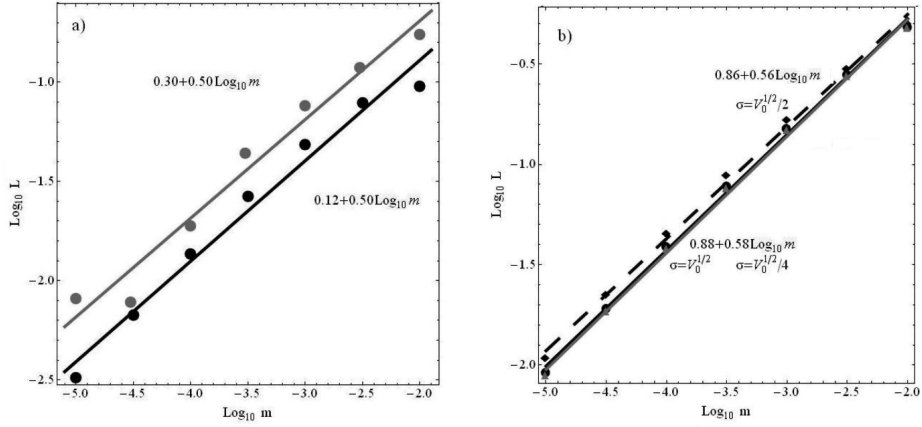}
\caption{(a)Mean finite Lyapunov characteristic number $L$ vs. the mass 
parameter $m$ as it follows theoretically from Eq.(\ref{finres}) (black dots), 
compared to the numerical computation (gray dots) at the integration time 
$t=10 ^{5}$, for all the experiments with $\varepsilon=0.1$. The mean 
inclination is close to 0.5 in the range $10^{-5}<m<10^{-2}$. 
(b) Theoretical computation of $L$ versus $m$ in a `isothermal' number 
density model (see text). The triangles and gray fitting line correspond 
to the choice $\sigma=V_0^{1/2}/4$, filled circles and black solid line 
to $\sigma=V_0^{1/2}/2$, squares and dashed line to $\sigma=V_0^{1/2}$.}
\label{finalincl}
\end{figure}
Assuming (as evidenced in Fig.\ref{liapt200}b) that a time $t=10^5$ is 
sufficient for a saturation of $\chi(t)$ close to the limiting value of 
chaotic orbits, i.e. close to the Luapunov characteristic number (LCN), 
the mean LCN of the orbits in an energy range $E_{min}\leq E\leq E_{max}$ 
can be estimated as 
\begin{equation}\label{meanlcn}
L\approx{1\over N_0}\int_{E_{min}}^{E_{max}} N(E)
\overline{\chi}(t=10^5,E)dE
\end{equation}
where $N_0=\int_{E_{min}}^{E_{max}} N(E)dE$, $N(E)$ is the number density 
of orbits of energy $E$, and $\overline{\chi}(t=10^5,E)$ is a mean estimate of 
the value of $\chi$ for orbits of energy $E$ at the final integration time. 
Considering only transiting orbits, we set $E_{min}=E_l\sim m^{2/3}$ 
(Eq.(\ref{qwer})). Also, from Fig.\ref{logElogllogm}a it is clear that the 
maximum energy $E_{max}=0.2$ considered in our samples is sufficiently high 
for the orbits' values of $\chi$ to fall with respect to the maximum by at 
least one order of magnitude in the worst case ($m=10^{-2}$), and typically 
by several orders of magnitude. 

The mean value $\overline{\chi}(t=10^5,E)$ can be now estimated by 
considering the separate mean values of $\chi$ as well as an $E$-dependent 
varying proportion of planar vs. 3D orbits. The percentage $\lambda$ of 
the planar orbits is estimated by the ratio of the surface occupied by initial 
conditions of planar orbits on the box surface corresponding to an energy 
level $E$ (see Fig.\ref{schematic}a) over the total area of the box. Thus 
\begin{equation}\label{}
\lambda={S_{2d}\over {S_{tot}}}\approx
{{48E^{1/2}r_m}\over{24E}}\approx{{2m^{1/3}}\over {E^{1/2}}}~~.
\end{equation}
The percentage of 3D orbits is $1-\lambda\approx {1-{2m^{1/3}}/{E^{1/2}}}$.
Finally, $\overline{\chi}$ is estimated according to Eqs.(\ref{lyap3d}), 
(\ref{lyap2d}), i.e.
\begin{equation}\label{meanchi23d}
\overline{\chi}_{planar}(E)\approx c_1{{m^{2/3}}\over E},~~~ 
\overline{\chi}_{3D}(E)\approx =c_2{m\over E^{3/2}}
\end{equation}
with $c_1$ and $c_2$ being constants of order unity. Then, for the mean value 
of $\chi$ at fixed energy  we have the estimate
\begin{equation}\label{chilam} 
\overline{\chi}(E)\approx\lambda\overline{\chi}_{planar}(E)
+(1-\lambda)\overline{\chi}_{3D}(E)~~.
\end{equation} 

The main uncertainty in Eq.(\ref{meanlcn}) regards the form of the 
number density function $N(E)$. In self-consistent models, $N(E)$ is determined 
by the distribution function of the centrophylic orbits. On the contrary, 
in `ad hoc' potential models, $N(E)$ cannot be determined self-consistently 
unless one possesses information on the kinematic distributions allowing to 
solve the reverse problem $density\rightarrow N(E)$. Assuming no detailed 
model, we hereby estimate the integral of Eq.(\ref{meanlcn}) using two 
different estimates of $N(E)$ as follows:

i) we consider the case of a nearly uniform distribution $N(E)=const$. 
Combining Eqs.(\ref{meanlcn}), (\ref{chilam}) and (\ref{meanchi23d}) 
we obtain:
\begin{equation}\label{finres}
L\approx{1\over {0.2-m^{2/3}}} 
{{[4c_1m^{2/3}+10c_2m^{4/3}-2.23(4c_1+2c_2)m]}}~~.
\end{equation}

The exact dependence of $L$ on $m$ in the model (\ref{finres}) depends 
on the relative values of the constants $c_1$ and $c_2$, as well as 
on factors entering in all the above estimates.  Figure \ref{finalincl}a 
shows $L$ against $m$ in logarithmic scale, by the estimate (\ref{finres}) 
(black points) setting simply $c_1=c_2=1$. The plot indicates an 
approximate power-law $L\sim m^p$. This is produced as follows: 
since $m$ is a small quantity, the leading term in Eq.(\ref{finres}) 
is the one with lowest exponent, i.e., $m^{2/3}$. Thus, in the absence 
of additional terms, we would have $p=2/3$. However, for relatively 
large $m$, the second most important term (linear in $m$) has a 
negative sign. Thus, it lowers the rightmost part of the curve $L$ 
vs. $m$ (the presence of $m^{2/3}$ in the denominator only marginally 
affects the overall power-law behavior). If we approximate the new 
curve by a single power-law fitting, we then find a lowering of the 
exponent $p$, i.e.
\begin{equation}\label{powlaw}
L\propto m^{2/3-q} 
\end{equation}
with $q$ varying between 0.1 and 0.2. However, one notices that the 
whole curve in log-log scale deviates downwards from the power-law 
approximation (\ref{powlaw}) for the highest mass parameter values, 
i.e. $q$ has a weak (increasing) dependence on increasing $m$.
Comparison with the numerical data (gray dots, for $\varepsilon=0.1$) 
shows that the model reproduces the slope of the numerical curve, which 
also exhibits a lowering of the value of $L$ at high mass parameter 
values with respect to an exact power law. Overall, the theoretical 
curve has a factor $\approx 2$ difference from the numerical curve, 
which is consistent with uncertainties in the theoretical coefficients. 
Notice also that a systematic lowering of the values of $L$ with respect 
to a power-law fitting is discernible in all the plots of Fig.\ref{plloglm}. 
Finally, we have checked that the appearance of an approximate power-law 
persists, with exponents around $p\approx 0.5$, for different choices 
of the constants $c_1$ and $c_2$. This shows that the dependence of 
the integral (\ref{meanlcn}) on $m$ (which enters in the integral 
as a parameter) is not sensitive on the details of the distribution 
of the planar vs. 3D orbits. 

ii) Fig.\ref{finalincl}b shows an evaluation of the integral (\ref{meanlcn}) 
for an isothermal (or `ergodic', see \cite{binntre2008}) model of $N(E)$, 
i.e. $N(E)\propto e^{-E/\sigma^2}$. The constant $\sigma$ is a 
measure of the velocity dispersion in the central parts of the galaxy. 
Assuming a core density $\rho\approx 3/(4\pi)$ (in our units, 
corresponding to total mass $M=1$ at radius $R=1$), by the Virial 
theorem $\sigma^2$ has to be taken of the order of the absolute value 
of the central potential well $V_0\approx \int_{0}^1 4\pi G\rho r dr 
\approx 1.5$. Figure \ref{finalincl}b shows the evaluation of the integral 
(\ref{meanlcn}) for three different choices of $\sigma$, namely 
$\sigma=V_0^{1/2}$, $\sigma=(1/2)V_0^{1/2}$ and $\sigma=(1/4)V_0^{1/2}$. 
In all three cases we recover here as well an effective power-law behavior, 
with not very different exponents, i.e. $p\simeq 0.58$, $p\simeq 0.56$ and 
$p\simeq 0.56$ respectively.

In conclusion, we find that an approximate power-law relation of the 
form (\ref{powlaw}) is robust against details of the form of the 
function $N(E)$.  
 
\section{Effect of central cusp}
In the model (\ref{vham}), the existence of many box orbits was a priori 
guaranteed due to the harmonic core in the center. It is well known, 
however, that realistic models of the central parts of galaxies include  
central density cusps $\rho(r)\sim r^{-\gamma}$ (see e.g. the review in 
\cite{binmer1998}, or \cite{mer1999}). In such models, the cusp itself 
transforms most centrophylic orbits to chaotic (\cite{mervalu1996}, 
\cite{merrquinl1998}). Even without central black hole, one then expects 
the orbits to exhibit positive Lyapunov exponents. We hereafter call this 
effect `residual chaos', i.e. chaos existing even when $m=0$. The corresponding 
mean Lyapunov exponent of the centrophylic orbits is denoted by $L_0$. 

\begin{figure}
\centering
\includegraphics[scale=0.4]{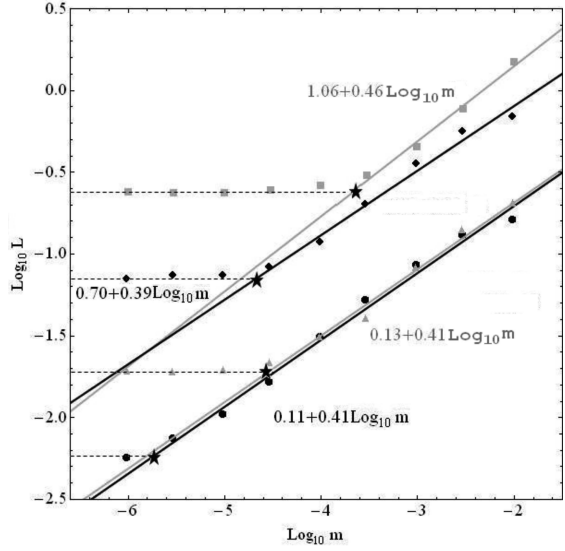}
\caption{The mean Lyapunov exponent versus the central mass parameter $m$ 
in four Dehnen triaxial models with central cusps (see text). Black filled 
circles correspond to the data points for the cusp exponent $\gamma=0.3$, 
gray triangles to $\gamma=0.7$, black rombuses to $\gamma=1.3$, and gray 
squares to $\gamma=1.7$. The straight lines (black or gray) are power law 
fittings obtained by the rightmost seven points for $\gamma=0.3$, five points 
for $\gamma=0.7$, five points for $\gamma=1.3$, and four points for $\gamma=1.7$. 
The horizontal dashed lines correspond the the values $\log_{10}L_0$, 
where $L_0$ is the `residual Lyapunov exponent' in each case (see text). 
The points where each horizontal line intersects the fitting line for 
the same corresponding $\gamma$ are marked as stars. }
\label{dehnen}
\end{figure}
Adding, now, a central mass ($m\neq 0$) we seek to determine the 
dependence of $L$ on $m$. The theoretical analysis of the previous section 
formally breaks down, since one cannot define the formal integrals 
$\Phi_x,\Phi_y,\Phi_z$ even for $m=0$. We thus rely on numerical 
computations. To this end, we consider again the Hamiltonian function 
(\ref{ham}), changing the potential model to
\begin{equation}\label{vcusp}
V=V_D-{m\over (r^2+d^2)^{1/2}} 
\end{equation}
where $V_D$ represents the ellipsoidal Dehnen model (\cite{den1993}):
\begin{equation}\label{vd}
V_D(x,y,z)=-\pi G a b c
\int_0^\infty {[\psi(\infty)-\psi(w)]d\tau\over
\sqrt{(\tau+a^2)(\tau+b^2)(\tau+c^2)}}
\end{equation}
where 
$$
\psi(w)=\int_0^{w^2}\rho(w'^2)dw'^2
$$
with
$$
w^2={x^2\over a^2}+{y^2\over b^2}+{z^2\over c^2},~~~a\geq b\geq c >0
$$
and $\rho(w)$ given by 
\begin{equation}\label{rhoden}
\rho(w)={(3-\gamma)M\over 4\pi a b c}
w^{-\gamma}(1+w)^{-4+\gamma}~~,~~~~0\leq\gamma<3~~. 
\end{equation}
The parameters $a$, $b$ and $c$ correspond to the lengths of the major, 
intermediate and minor axis of the triaxial equipotential surface 
corresponding to $w=1$. The parameter $M$ determines the system's total 
mass. We use a similar algorithm as in \cite{merrfrid1996} in order to 
numerically evaluate the integral (\ref{vd}) as well as its spatial 
derivatives, i.e., the forces. 

A `weak cusp' corresponds to $\gamma<1$. In this case, the modulus of 
any component of the force generated by $V_D$ goes to zero at the center, 
reaches a maximum value at a certain distance from the center, and then 
falls-off tending to zero at large distances by a Keplerian law. This 
behavior of the force allows to determine values of the parameters 
$a$, $b$, $c$, and $M$, for fixed $\gamma$, so as to create a system 
exhibiting a similar geometry and value of the total mass as the 
simplified system corresponding to the potential (\ref{vham}) of 
section 2, the two systems being, hence, differentiated essentially 
only by the presence of a central cusp as opposed to a harmonic core 
respectively. The parameter determination is realized by the following 
algorithm: 

i) We set the ratios $a:b = \omega_y:\omega_x$ and $a:c = \omega_z/\omega_x$, 
where $\omega_x$, $\omega_y$, and $\omega_z$ are the parameters of 
(\ref{vham}). 

ii) We fix the value of $a$ so that the quantity $\partial V_D/\partial x$ 
presents maximum at the point $(x=x_{max},y=0,z=0)$, with $x_{max}$ 
choosen so as to represent the point where the harmonic model in 
(\ref{vham}) yields a total mass equal to unity. We find $x_{max}
\simeq 1.06$. 

iii) We fix $M$ so that the force $F_x$ under the potential $V_D$ be 
equal to the force $F_x$ under the potential (\ref{vham}), with $m=0$, 
at the point $(x=x_{max},y=0,z=0)$. Note that, since the value of 
the force depends essentially only on the total mass inside a 
given radius, this normalization means also that the total mass 
inside an ellipsoidal surface crossing $x=x_{max}$ is nearly equal 
in the harmonic and in the $\gamma$--models.  

We examine two values of $\gamma$ in the weak cusp case, namely 
$\gamma=0.3$ and $\gamma=0.7$.

In the case, now, of a `strong cusp' ($\gamma>1$), criterion (ii) can 
no longer be implemented, since $\partial V_D/\partial x\rightarrow\infty$ 
as $x\rightarrow 0$ (with $y=z=0$), implying that the quantity 
$\partial V_D/\partial x$ does not present any smooth maximum along any 
of the principal axes. As a simple (but somewhat arbitrary) way to bypass 
this difficulty, we keep $a$ constant to the value $a=6.67$ found by 
criterion (ii) in the second `weak cusp' experiment ($\gamma=0.7$). 
Then, we fix the remaining constants by criteria (i) and (iii). 
We run also two strong cusp experiments, with $\gamma=1.3$ and 
$\gamma=1.7$. 

In all four experiments, the initial conditions are choosen as in section 
2, namely 200 initial points of zero velocity randomly distributed on 
equipotential surfaces with $V=E$ and $E$ choosen uniformly in the range 
$0\leq V\leq E_{max}$, with $E_{max}$ choosen as $E_{max}=V_D(x_{max})$, 
so as to ensure that the resulting centrophylic orbits exhibit oscillations 
of amplitude at most equal to $x_{max}$. However, here we integrate the 
orbits only up to $t=1000$, since the complexity of force evaluation in 
the model $V_D$ renders the computational cost of longer integration 
prohibitive. Yet, as shown below, our smallest found Lyapunov exponents 
are about $L\approx 10^{-2.5}$, implying that a time $t=10^3$ is marginally 
greater than the saturation time $t\sim 1/L$ even for the orbits 
with smallest Lyapunov exponents.  

Figure \ref{dehnen} (analogous to Fig.\ref{plloglm}, top row) shows the 
mean Lyapunov number $L=\overline{\chi}(t=10^3)$ for our ensembles of orbits 
in the four above experiments, as a function of the central mass parameter 
$m$. We note immediately that power-law fittings are possible in only 
a range of values of $m$, i.e. above a critical threshold value 
$m>m_c(\gamma)$. In Fig.\ref{dehnen}, an estimate of the threshold 
value is found by the abcissa of the points displayed by stars. They are 
computed as follows: the four inclined lines represent power-law fittings 
for the rightmost part of the numerical curve of $L$ vs. $m$ in each 
experiment. The horizontal lines illustrate the level values of the 
quantity $L_0=L(m=0)$. We call $L_0$ the `residual Lyapunov exponent'. 
It represents the mean Lyapunov exponent of the centrophylic orbits 
when $m=0$, i.e., under the influence of the central cusp only. 
The point at which a horizontal line of fixed $L_0$ intersects 
the corresponding inclined fitting line of $L$ vs. $m$ in the same 
experiment (same $\gamma$) marks the position of a star-point, and the 
associated abscissa, i.e. a critical mass value $m_c$.   From 
Fig.\ref{dehnen} it is straightforward to see that both  
$L_0$ and $m_c$ increase in general with the strength of the cusp 
(i.e. the value of $\gamma$). On the other hand, it is clear from 
the numerical data points that an approximate power-law correlation 
between $L$ and $m$ persists, in all four experiments, for central 
mass parameters larger than $m=m_c$. A physical understanding of this 
phenomenon is the following: the chaotic scattering caused by the cusp 
itself acts dynamically as a central mass concentration, whose distribution 
is not point-like but follows the cusp density law. As long as the BH mass 
is small, the effect of the cusp is dominant over the efect of the BH. 
Thus, the mean Lyapunov exponent of the centrophilic orbits remains 
nearly equal to the residual Lyapunov exponent $L_0$. But beyond the 
BH mass scale $m=m_c$, the black hole dominates over the central 
cusp. Then, we recover a correlation of $L$ with $m$. This goes 
asymptotically to an effective power-law. Furthermore, the exponents 
found by fitting in the range $m>m_c$ are all about $p\simeq 0.4$, i.e., 
not very different from those of the corresponding data in Fig.\ref{plloglm}, 
the essential difference in the two plots being with respect to whether 
or not we observe a a critical mass scale in which the power-law breaks.   

We note finally, that despite the sparsity of their datapoints, the results 
of \cite{mervalu1996}, reproduced here as Fig.\ref{mervalkand}b, show 
essentially the same structure as the results of Fig.\ref{dehnen}. 
Thus, the residual chaos phenomenon explains the plateau of the curve 
$L$ vs. $m$ in the data of \cite{mervalu1996} as well. 
 
\section{The $L\propto m^p$ law in disc-barred galaxies}

N-body simulations of barred galaxies (e.g. \cite{frieben1993}, \cite{frie1994}, 
\cite{normshellhas1996})  have demonstrated that the growth of a central 
mass concentration induces secular evolution in such systems as well. In fact, 
although dynamically not favored, black hole growth to a mass level as high 
as $10^8M_\odot$--$10^9M_\odot$, corresponding to a few percent of the mass 
of a typical galactic bar, could induce even a total destruction of the bar, 
with its conversion into a nearly axisymmetric bulge-like component. 
Test particle integrations in barred potentials (\cite{pfenn1984}, 
\cite{pfenndezeu1989}, \cite{hasetal1993}), \cite{normshellhas1996}, 
\cite {shen2004}, indicate that a primary mechanism responsible for 
the secular evolution of bars, and even bar dissolution, is chaos 
induced by the central mass. 

Hereafter we study the dependence of Lyapunov exponents on the central mass 
paramerer in rotating disc-barred galaxies. Two points should be immediately 
emphasized: i) our modeling in previous sections was based on the existence 
of box-like centrophilic chaotic orbits. Such orbits cannot exist in rotating 
disc-barred galaxies. However, as shown in the sequel, centrophilic orbits 
appear around the main families of planar periodic orbits (e.g. the $x_1$ 
family). Note that the presence of some type of centrophilic orbits 
is an indispensible feature of bars with a rising density profile in the 
center. As discussed below, albeit different in morphology, the centrophilic 
orbits in barred galaxies are found numerically to exhibit a similar chaotic 
behavior as the boxy centrophilic orbits in elliptical galaxies. ii) 
Besides the central mass, chaos is generated by the interaction of resonances 
in the corotation domain 
(\cite{con1981}, \cite{pfenn1984}, \cite{spasel1987}, \cite{pfefri1991}, 
\cite{kaufcon1996}, \cite{patetal1997}, \cite{fux2001}, \cite{pichetal2004}, 
\cite{kaupat2005}). Nevertheless, this type of chaos is a quite distinct 
phenomenon. In fact, most chaotic orbits in the corotation domain belong 
to the so-called `hot population' (\cite{spasel1987}), hence they are not 
centrophilic. 

\subsection{Potential model}
As a case study, we employ the barred-galaxy potential introduced by 
\cite{kaufcon1996} in a rough self-consistent modelling of the 
galaxy NGC3992. Adding a component for the central mass, the total potential 
is analyzed as:
\begin{equation}\label{potall}
V_{tot}=V_{bh} + V_{h} + V_{d} + V_b
\end{equation}
where $V_{bh}$ is the potential generated from the central mass (black hole), 
while $V_{h}$, $V_{d}$, $V_b$ are dark halo, disc, and bar potential components 
respectively.  The potential of the central mass is, as before, 
\begin{equation}
V_{bh}=-{m_{bh}\over{\sqrt{r^2+d^2}}}~~.
\end{equation}
The remaining terms are as in \cite{kaufcon1996}. The dark halo term is 
\begin{equation}\label{poth}
V_{h}(r)={{-M_{h}}\over{\sqrt{r^2+b_h^2}}}~~.
\end{equation}
The disc term corresponds to an exponential disc:
\begin{equation}
V_{d}(r)=-\Sigma_0\pi r[I_0({1\over2}\epsilon_d
r)K_1({1\over2}\epsilon_d r)-I_1({1\over2}\epsilon_d
r)K_0({1\over2}\epsilon_d r)]
\end{equation}
where $I_0,$ $I_1$ and $K_0$, $K_1$ are modified bessel functions of the 
first and second kind respectively. The bar term is of the Ferrers $n=2$ 
type, with the major axis alligned with the y-axis:
\begin{eqnarray}
V_{b}(x,y,z)&=&\nonumber-{105M_b\over
96}[3(2W_{110}x^2y^2-W_{120}x^4y^2-W_{210}x^2y^2-W_{100}y^2+W_{020}x^4+W_{200}y^4
\\&~& -W_{010}x^2)+W_{000}-W_{030}x^6-W_{300}y^6]
\end{eqnarray}
where the coefficients $W_{ijk}$ are given by elliptic integrals. 
All model's parameters, as well as the value of the pattern speed $\Omega_p$ 
are as in \cite{kaufcon1996}, referring to the model for the galaxy NGC3992. 
They are summarized in table I below. Note that the original model contains 
also a spiral-arm term, which, however, is only important at radii beyond 
the end of the bar, and it is here ignored. 
\begin{table}\label{I}
\centering
\begin{tabular}{|c c c c c c c c c|}
\hline\hline 
$ Bar:$  & $M_b$ & $\alpha$ & $b$ & $c$ & $\Omega_p$ &  & &  \\
$ $  & $1.5$ & $5.5$ & $2.1$ & $0.55$ & $43.6$ & $$ & $$ & $$ \\
\hline
$Disc: $  & $\Sigma_0$ & $\varepsilon_d$ &  &  &  &  &  &  \\
$ $  & $750$ & $0.235$ &  &  &  &  & &  \\
\hline
$Halo: $  & $M_h$ & $b_h$ &  &  &  &  &  &  \\
$$  & $27.5$ & $12$ &  &  &  &  &  &  \\ [1ex] 
\hline 
\end{tabular}
\caption{Parameters of the disc-barred galactic potential. The units are 
$Kpc^{-1}$ for $\varepsilon_d$, $Kpc$ for $a$, $b$, $c$ and $b_h$, 
$10^{10} M_\odot$ for $M_h$, $M_b$, $m_{bh}$, $Km s^{-1} Kpc^{-1}$ for 
$\Omega_p$, $M_\odot/pc^2$ for $\Sigma_0$.}
\end{table}

\subsection{Numerical experiments}
As in section 2, we numerically integrate the equations of motion, as well as 
the variational equations, for planar orbits under the Hamiltonian (in cylindrical 
coordinates):
\begin{equation}\label{hamax}
H(r,\theta,p_{r},p_{\theta})=
{1\over2}{(p_{r}^2+{p_{\theta}^2\over{r^2}})}-\Omega_p p_\theta+V(r,\theta)=E_j
\end{equation}
where $E_j$ is the jacobi constant.
The Hamiltonian (\ref{hamax}) describes the motion in a rotating frame with 
pattern speed $\Omega_p$, while $p_{\theta}$ is the angular momentum 
in the inertial frame of reference.

Initial conditions are choosen in a way so as to ensure that they give rise 
to centrophilic orbits. To this end, we consider, as above, ensembles of 200 
orbits with initial positions uniformly distributed on a cycle of radius 
$r=0.1$ around the galactic center. Initial velocities are given in the 
direction radially outwards, with modulus choosen so that the value of 
the Jacobi constant is uniformly distributed in the range $-2.16\times 
10^{5}\leq E_j\leq -2.03\times 10^{5}$. This range is choosen so as to 
correspond to energies well below the value at the Lagrangian equilibrium 
point $L_1$, i.e. $E_{j,1}=-1.915\times 10^5$. The corresponding orbits 
lie then always inside the corrotation domain, i.e. they support the bar. 
Orbit ensemble integrations are done for a time $t=10^{5}$ (in comparison, 
orbital periods are of order $\sim 0.1$). In different experiments the 
central mass varies in the range $10^{3}M_{\odot}\leq m_{bh}\leq 
10^{7}M_{\odot}$.

\begin{figure}
\centering
\includegraphics[scale=0.6]{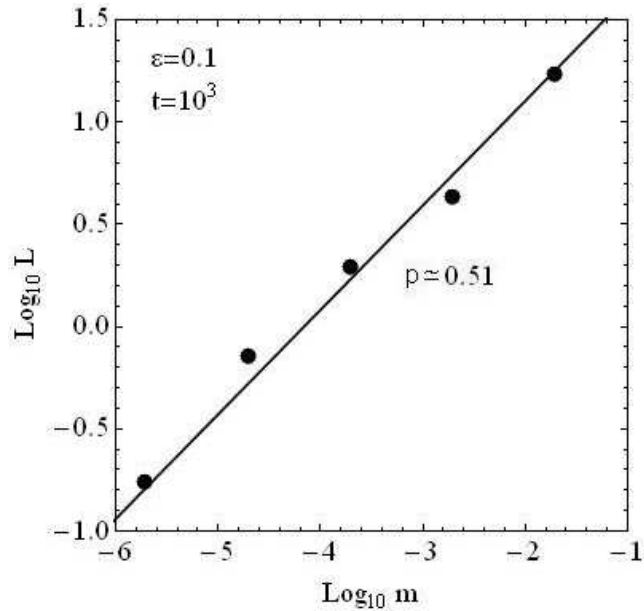}
\caption{Logarithm of the mean Lyapunov number of the orbits in the 
ensembles of initial conditions as described in the text versus the 
central mass parameter $m=m_{bh}/m_{bar}$ (in logarithmic scale) in 
a disc-barred galaxy model. The logarithmic slope indicates a power-law 
exponent $p=0.51$.} 
\label{plloglmspir}
\end{figure}
Figure \ref{plloglmspir} shows the main result: the mean Lyapunov number $L$
of the chaotic orbits choosen as above is plotted against the mass 
parameter $m$, choosen here as the ratio $m=m_{bh}/m_{bar}$, since this 
ratio is relevant to a quantification of the rate of secular evolution 
of the bar. We observe again that the numerical data can be fitted by 
a power-law $L\propto m^p$, with $p=0.51$. 

We now interpret the mechanisms of chaos and identify the families of orbits 
which are responsible for this behavior. 

\subsection{Interpretation}
The mechanism by which a central black hole generates chaos in a disc-barred 
galaxy can be visualized with the help of phase portraits, obtained by means 
of a suitable surface of section. Here we employ the {\it apocentric} condition 
$\dot{r}=0$, $\ddot{r}<0$, in order to define the surface of section. We then 
plot the intersection points of all orbits with the above section as projected 
on the plane $(\theta,p_\theta)$. 

\begin{figure}
\centering
\includegraphics[scale=0.8]{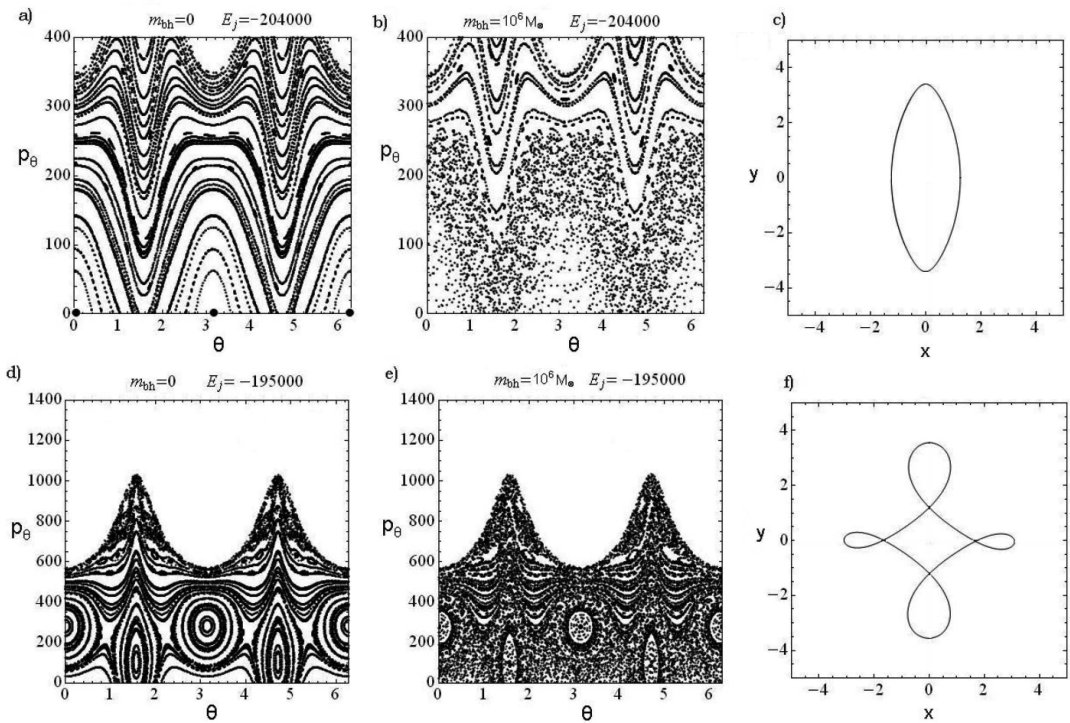}
\caption{ Apocentric Poincar\'{e} surfaces of section with $m_{bh}=0$ at the 
energies (a) $E_J=-204000$ and (d) $E_J=-195000$. (b,e) Same as in (a,d), but 
now with a central black hole of mass $m_{bh}=10^{6}M_{\odot}$. The $x_1$ 
periodic orbit (c) at the energy $E_J=-204000$ has a 2:1 form (two apocentric 
passages per revolution in the rotating frame), while (f) at the energy 
$E_J=-195000$ it has a 4:1 form.} 
\label{poinc2to14to1}
\end{figure}

\begin{figure}
\centering
\includegraphics[scale=0.50]{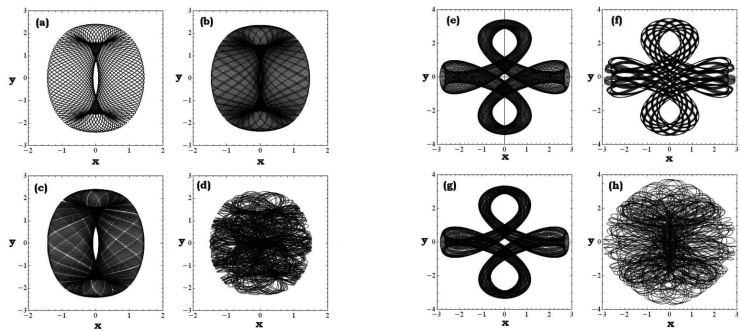}
\caption{(a) Tube orbit with energy -216000 around the 2:1 periodic orbit 
before the insertion of the central mass.
The initial conditions correspond to a librational invariant curve of the 
2:1 island of stability which survives after the insertion of the black hole.
(b) Same as in (a), but for initial conditions on an invariant curve of the 
2:1 island of stability which is destroyed after the insertion of the black 
hole. (c) The orbit with same initial conditions as in (a) remains ordered 
after the insertion of a central mass $m_{bh}=10^{6}M_{\odot}$. 
(d) On the contrary, the orbit with same initial conditions as in (b) 
is converted to chaotic. (e,f) Same as in (a,b) but for two tube orbits 
of the 4:1 resonance. (g,h) The orbits with the same initial conditions as 
in (e,f) respectively, after the insertion of the central mass 
$m_{bh}=10^{6}M_{\odot}$. The one of (g) remains regular, while the one 
of (h) is chaotic. } 
\label{orb2to14to1}
\end{figure}

Figure \ref{poinc2to14to1} shows the surface of section portrait at the 
energies (Jacobi constant values) $E_j=-204000$ and  $E_j=-195000$, without 
central black hole (Fig.\ref{poinc2to14to1}a,d), or with a black hole of mass 
$m_{bh}=10^{6}M_{\odot}$ (Fig.\ref{poinc2to14to1}b,e). The change of phase space 
structure is evident, namely the insertion of the central mass destroys 
many rotational KAM curves, corresponding to regular (quasi-periodic) 
orbits around the galactic center. Also, at the second energy level 
(lower panels), which corresponds to motion closer to corotation, 
a number of librational KAM curves around the 1:4 island of stability 
are destroyed. In fact, one can see that at both energy levels the 
orbits converted to chaotic are {\it tube} orbits around the 2:1 and 
4:1 branches of the $x_1$ stable periodic orbit. The periodic orbits 
themselves at the corresponding energies are shown in 
Fig.\ref{poinc2to14to1}c and f respectively. 
The 2:1 orbits exist up to energies $\approx-199000$. When $m_{bh}=0$, 
the quasi-periodic tube orbits around a 2:1 orbit give rise to two islands 
of stability in the surface of section. The outermost librational invariant 
curves of these islands correspond to thick tube orbits (Fig.\ref{orb2to14to1}). 
The crucial difference between the two tube orbits in Figs.\ref{orb2to14to1}a 
and b is that in Fig.\ref{orb2to14to1}a the tube orbit leaves a hole in the 
center, whose dimension is larger than the black hole's sphere of influence. 
On the contrary, the tube thickness of the orbit in Fig.\ref{orb2to14to1}b 
is larger than the half-width of the orbit's amplitude of oscilation in the 
y-axis, i.e., the orbits leaves no hole in the center. Then, after the insertion 
of the central mass, the orbit with same initial conditions as in 
Fig.\ref{orb2to14to1}a retains its regular (quasi-periodic) character 
(Fig.\ref{orb2to14to1}c), while the orbit with same initial conditions 
as in Fig.\ref{orb2to14to1}c becomes chaotic (Fig.\ref{orb2to14to1}d). 
A similar criterion applies to whether or not a thick tube orbit around 
the 4:1 periodic orbit becomes regular or chaotic after the insertion of 
the central mass (Fig.\ref{orb2to14to1}e-h). In fact, one readily finds 
that the initial conditions separating these two types of orbits correspond 
to the last librational KAM curve in the 4:1 island of stability of the 
section of Fig.\ref{poinc2to14to1}d.  

Investigating the efficiency of the above mechanism to other resonances, one 
finds that the mechanism is not able to produce chaotic orbits in the cases 
of other low-order resonances like 3 : 1 and 6 : 1. In fact, the tube orbits 
trapped around these central periodic orbits in this model form rather small 
islands of stability. Thus, the tube thickness is small, and we find no 
tube orbits able to cross the central masses' sphere of influence. A 
similar restriction holds for higher order families in the same model. 

\section{Conclusions}

In the present paper we analyze the origin of a numerically observed  
approximate power-law relation $L\propto m^p$, with $p\sim$0.3--0.5, where 
$L$ is the mean Lyapunov exponent of centrophilic orbits in galaxies with 
central masses (black holes), and $m$ the mass parameter, i.e., the ratio 
of the central mass over the mass of the galaxy. Also, we find that 
such a law can be recovered in quite different contexts and models of 
galactic systems, ranging from elliptical galaxies with cores or cusps 
to rotating barred galaxies. In particular:

i) We first make numerical experiments with a simple model of elliptical 
galaxy with smooth central force field, to which the force field of the 
central mass is superposed. The experiments confirm the power-law $L\propto m^p$, 
when $L$ is estimated through its `finite time' analog, i.e. the mean value 
of finite-time Lyapunov exponents. We demonstrate the statistics of these 
values for centrophilic orbits. We also find that $p$ has a tendency towards 
the upper limit 0.5 at longer integration times.

ii) We demonstrate that the law $L\propto m^p$ can be extracted also from 
compiling data of previous works in the literature (\cite{mervalu1996}, 
\cite{kandrsid2002}), in galaxies with both smooth and cuspy centers. 

iii) We make a theoretical analysis of the Lyapunov exponents for centrophilic 
box-like orbits in elliptical galaxies. We demonstrate that the mean Lyapunov 
exponent can be obtained by independently estimating a) the mean value of the 
so-called `stretching number' (=one-step Lyapunov number) at every transit of 
an orbit from the sphere of influence of the central mass, and b) the rate 
of visits of the orbits to the sphere of influence. In both cases, we find 
how the various estimates depend on $m$ as well as on the orbital energy. 
Regarding (b), we find two different estimates, according to whether an orbit 
can be characterized as `planar' or `3D'. Putting all estimates together, 
one arrives at a theoretical reproduction of the $L\propto m^p$ law. 

iv) In the case of models with central cusps, we find a critical mass 
scale $m_c$, such that for central mass parameters $m<m_c$ the chaotic 
behavior of the centrophylic orbits is dominated by the central cusp 
(we call this `residual chaos'), while for $m>m_c$ an approximate 
power-law correlation $L\propto m^p$ is restored, with $p\approx 0.4$. 
The critical mass scale $m_c$ as well as the `residual mean Lyapunov 
exponent' $L_0$ are increasing functions of the exponent $\gamma$ in 
power-law central cusps $\rho(r)\sim r^{-\gamma}$.

v) We finally explore numerically the correlation between $L$ and 
$m$ for the centrophylic orbits in disc galaxies with rotating bars. 
In this case, while there can be no box-like centrophilic orbits, 
we find several quasi-periodic tube orbits around the main families 
of periodic orbits (like 2:1 or 4:1), for which the tube is thick 
enough so as to pass arbitrarily close to the center. These orbits 
support the rising density profile of the bar at the center. Their 
initial conditions are close to the last librational KAM curve of 
the islands of stability around their corresponding periodic orbits. 
Numerically, we observe that only the tube orbits around the 
lowest-order periodic orbits can become centrophylic. Furthermore, 
for the chaotic counterparts of these orbits, after the insertion of 
the black hole, we numerically recover again a correlation of the 
form $L\propto m^p$. 

\section*{Acknowledgments}
This research is supported by the Research Committee of the Academy of 
Athens (grant 200/815). N. Delis was supported by the State Scholarship 
Foundation of Greece (IKY).

\clearpage
\noindent
{\bf Appendix}

\begin{figure}
\centering
\includegraphics[scale=0.50]{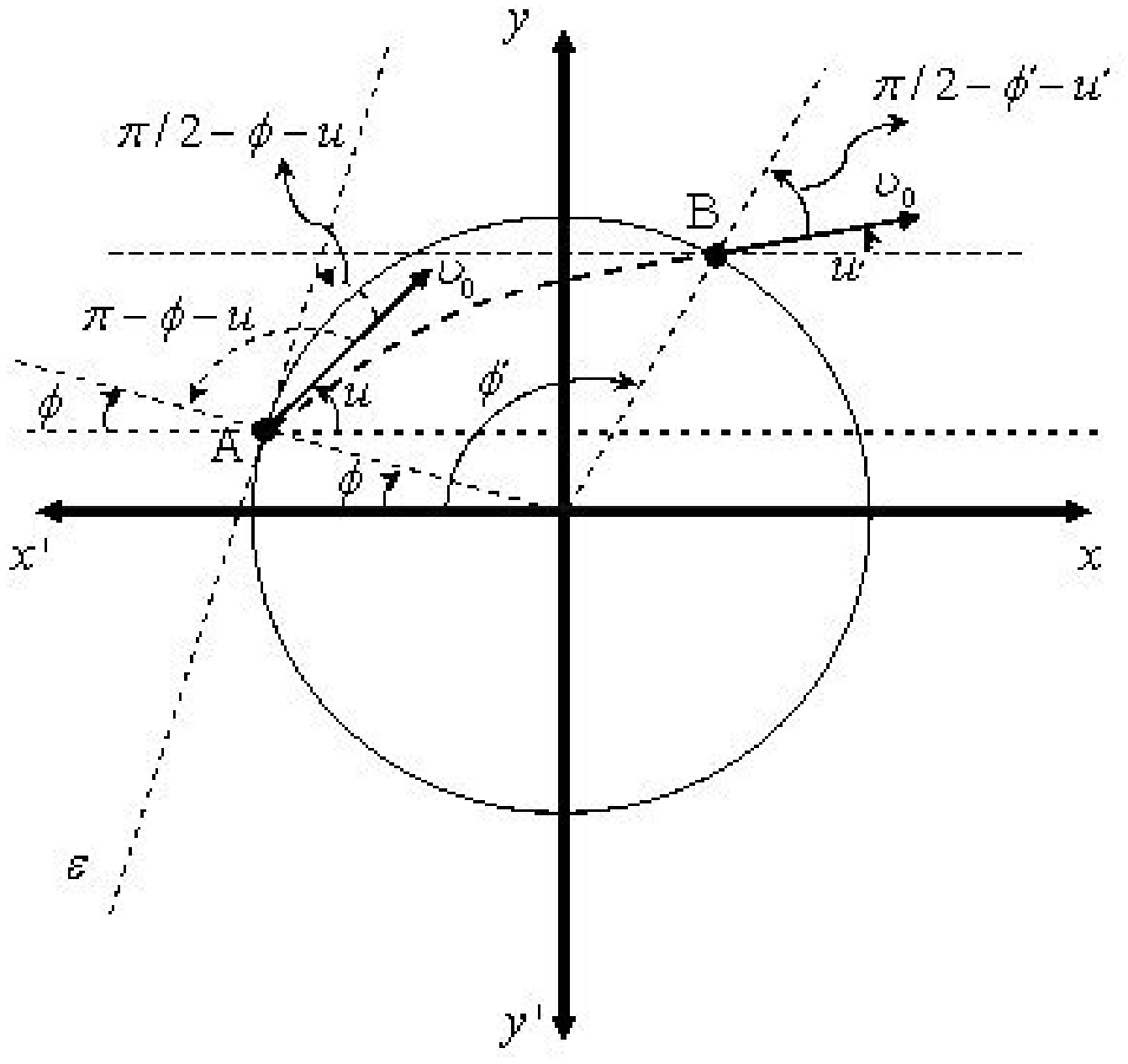}
\caption{Schematic representation of our model for orbit transits 
via the sphere of influence of the central mass. An orbit enters the 
sphere at point A and exits at point B. Hyperbolic Keplerian dynamics 
is assumed in order to estimate the orbit's local `stretching number' 
arising at the transit (see text).} 
\label{2Dkeplerab}
\end{figure}
We theoretically estimate the local value of the `stretching number' of an 
orbit transiting the sphere of influence of the central mass (Fig.\ref{2Dkeplerab}), 
schematic, as follows: approximating the motion during the transit as Keplerian, 
in polar co-ordinates $(r,\phi)$ with respect to the origin on a plane including 
the origin and the entry and exit points (A,B respectively), one has
\begin{equation}\label{kepler}
{1\over r}={{Gm\over C^{2}}+A\cos(\phi-\phi_0)}
\end{equation}
where $C$ is the local value of the angular momentum (assumed constant during 
the transit, see section 3). The constant $A$ is given by:
\begin{equation}\label{}
A=e{Gm\over C^{2}},
\end{equation}
where $e=\sqrt{1+{2EC^2}/{mG^2}}$, whereas $\phi_0$ is the angle corresponding 
to the closest approach to the origin, at distance ($r_{min}$. The orbit has the 
the same velocity measure $v_0$ at the points A and B. We assume that the velocity 
vector at A forms an angle $u$ with the horizontal axis. 

With the above conventions, to a given entry angle $\phi$ corresponds a given 
exit angle $\phi'$ from the sphere of influence. After some algebra (taking 
into account Eq.(\ref{kepler}) as well as the preservation of energy and 
angular momentum), we find:
\begin{eqnarray}\label{df2}
{\phi-\phi'}&=&2\cos^{-1}\nonumber
[{{\upsilon_0\sin{(\phi+u)}}\over{\sqrt{{({Gm/ {{r_m \upsilon_0
\sin{(\phi+u)}}
}})}^2+\upsilon_0^2-{2Gm/{r_m}}}}}
\\&~~~~~&{({1-{Gm\over{{{r_m\upsilon_0^2 \sin^2{(\phi+u)}}}}}})}]~~.
\end{eqnarray}

The local value of the stretching number can now be estimated by the difference 
in $\phi'$ for two nearby orbits entering at slightly different angles $\phi$. 
Taking the derivative of eq.(\ref{df2}) we have:
\begin{eqnarray}\label{dfdf}
|{d\phi'\over{d\phi}}|=|1+{d\over{d\phi'}}[2\cos^{-1}&[&\nonumber
{{\upsilon_0\sin{(\phi+u)}}\over{\sqrt{{({Gm/ {{r_m \upsilon_0
\sin{(\phi+u)}}
}})}^2+\upsilon_0^2-{2Gm/{r_m}}}}}\\&~&{({1-{Gm\over{{{r_m
\upsilon_0^2 \sin^2{(\phi+u)}}}}}})}]]|~~.
\end{eqnarray}
Re-orienting the frame of reference, without loss of generality the parameter 
$u$ can be set equal to zero. The quantity $|{d\phi'/{d\phi}}|$ is the measure 
of the stretching number $a$ for the transit. motion inside the sphere $r_m$.
One can check that for typical energies of centrophilic orbits, one has 
$2mr_m\upsilon_0^2<<r_m^2\upsilon_0^4$. Then, Eq.(\ref{dfdf}) reduces to:
\begin{equation}\label{dfdf3}
|{d\phi'\over{d\phi}}|\simeq|{1-2{{1+{{m\csc^2\phi}/{r_m\upsilon_0^2}}}
\over{1+{{m^2\csc^2\phi}/{r_m^2\upsilon_0^4}}}}}|~~.
\end{equation}
The mean stretching number of a transit at given velocity $\upsilon_0$, with 
respect to all possible angles $\phi$ can be estimated by the integral
of the quantity $\log|{d\phi'/{d\phi}}|$ over all possible angles
\begin{eqnarray}\label{lyap}
\overline{\log\left|{d\phi'\over{d\phi}}\right|}&=&\nonumber
{2\over{\pi}}\int_{0}^{\pi/2}\log|{d\phi'\over{d\phi}}|d\phi=
\\&~&{2\over{\pi}}\int_{0}^{\pi/2}\log|{1-2{{1+{m\over{r_m\upsilon_0^2}}\csc^2\phi}
\over{1+{m\over{r_m^2\upsilon_0^4}}\csc^2\phi}}}|d\phi
={2\over{\pi}}\int_{0}^{\pi/2}\log|{(1-2{{\alpha_0+\sin^2\phi}\over{\alpha_0^2+\sin^2\phi}})|}d\phi=
\\&~&\nonumber{2\over{\pi}}\int_{0}^{\pi/2}\log{(2{{\alpha_0+\sin^2\phi}\over{\alpha_0^2+\sin^2\phi}}-1)}d\phi
=\log{(2{{1+\alpha_0}\over{1+\alpha_0^2}}-1)}+
2\log{\sqrt{1+\alpha_0^2}(1+\sqrt{{\alpha_0(\alpha_0-2)}
\over{\alpha_0^2-2\alpha_0-1}})\over{\alpha_0+\sqrt{1+\alpha_0^2}}}
\end{eqnarray}
where $\alpha_0=m/r_m\upsilon_0^2$. At mass ranges $10^{-5}<m<10^{-2}$ the value 
of $\alpha_0$  is in general a small quantity, except for small energies ($E<0.03$ 
in our units), which, however, can be readily checked to correspond to orbits 
residing always within the sphere $r=r_m$, i.e. pyramids (\cite{merrvasil2011}). 
Excluding such orbits, we set $\alpha_0^2=0$ in the previous equation and obtain:
\begin{eqnarray}\label{lyap2}
\overline{\log\left|{d\phi'\over{d\phi}}\right|}\simeq\log{({1+2\alpha_0})}+
2\log{(1+\sqrt{{-2\alpha_0}\over{-2\alpha_0-1}})\over{a+1}}\nonumber \\
= \log{({1+2\alpha_0})}+
2\log{(1+{{\sqrt{\alpha_0}}{\sqrt{{-2}\over{-2\alpha_0-1}}}})\over{a+1}}~~.
\end{eqnarray}
Expanding Eq.(\ref{lyap2}) with respect $a_0^{1/2}$ we find:
\begin{eqnarray}\label{dfdf4}
\overline{\log|{d\phi'\over{d\phi}}}|\simeq
2{\alpha_0}+2\log{(1+{\sqrt{\alpha_0}{\sqrt{2}}})}
=2{\alpha_0}+2{\sqrt{2}\sqrt{\alpha_0}}\nonumber \\
={2\over{(r_m\upsilon_0^2)}}m+{\sqrt{2}2\over{(r_m\upsilon_0^2)^{1/2}}}\sqrt{m}
\simeq{\sqrt{8\over{(r_m\upsilon_0^2)}}}\sqrt{m}~~.
\end{eqnarray}
In Eq.(\ref{dfdf4}) the quantity $r_m$ depends on $m$ ($r_m\propto m^{1/3}$).
Furthermore in the limit of the sphere of black hole's influence, energy
is approximately equal with the kinetic energy ($\upsilon_0^2\propto E$), 
whereby the last expression can be written in the form:
\begin{equation}\label{dfdf5}
\overline{a}\approx\overline{\log|{d\phi'\over{d\phi}}}|\propto {m^{1/3}\over E^{1/2}}~
\end{equation}
i.e. we arrive at the estimate of Eq.(\ref{meana}). 


\begin{thebibliography}{}
\bibitem[\protect\citeauthoryear{Binney \& Merrifield}{1998}]{binmer1998} 
Binney, J., Merrifield, S., 1998, `Galactic Astronomy', Princeton 
University Press.

\bibitem[\protect\citeauthoryear{Binney \& Tremaine}{2008}]{binntre2008} 
Binney, J., Tremaine, S., 2008, `Galactic Dynamics', 2nd edition, Princeton 
University Press.

\bibitem[\protect\citeauthoryear{Contopoulos}{1960}]{cont1960} 
Contopoulos, G., 1960, Zeitschrift fur Astrophysik, 49, 273

\bibitem[\protect\citeauthoryear{Contopoulos}{1981}]{con1981} 
Contopoulos, G., 1981, Astron. Astrophys., 102, 265

\bibitem[\protect\citeauthoryear{Contopoulos et al.}{2002}]{convogcal2002} 
Contopoulos, G., Voglis, N., Kalapotharakos, C., 2002, Cel. Mech. Dyn. Astron., 
83,  191

\bibitem[\protect\citeauthoryear{Dehnen}{1993}]{den1993} 
Dehnen, W., 1993, Mon. Not. R. Astron. Soc., 265, 250.

\bibitem[\protect\citeauthoryear{Debattista}{2006}]{Debat2006} 
Debattista, V. P., 2006, New Horizons in Astronomy: Frank N. Bash 
Symposium, ASP Conference Series, 352

\bibitem[\protect\citeauthoryear{Efthymiopoulos et al.}{2004}]{eftetal2004}
Efthymiopoulos, C., Contopoulos, G., Giorgilli, A., 2004, J. Phys. A Math 
Gen, 37, 10831. 

\bibitem[\protect\citeauthoryear{Efthymiopoulos et al.}{2007}]{efthvogkal2007}
Efthymiopoulos, C., Voglis, N., Kalapotharakos, C., 2007, Lect. Notes 
Phys. 729, 297

\bibitem[\protect\citeauthoryear{Faber et al.}{1997}]{faberetal1997}
Faber, S. M., Tremaine, S., Ajhar, E.A., Byun, Y., Dressler, A., 
Gebhardt, K., Grillmair, C., Kormendy, J., Lauer, T.R., Richstone, D., 
1997, Astron. J., 114, 1771

\bibitem[\protect\citeauthoryear{Ferrarese et al.}{1994}]{feretal1994}
Ferrarese, L.,van den Bosch, F.C., Ford, H.C., Jaffe, W., O'Connell, R.W.,
1994, Astron. J., 108, 1598. 

\bibitem[\protect\citeauthoryear{Ferrarese \& Ford}{2005}]{ferrford2005}
Ferrarese, L., Ford, H.C., 2005, Space Sci. Rev., 116, 523

\bibitem[\protect\citeauthoryear{Fridman \& Merritt}{1997}]{fridmanmer1997}
Fridman, T., Merritt, D., 1997, Astron. J., 114, 1479

\bibitem[\protect\citeauthoryear{Friedli \& Benz}{1993}]{frieben1993}
Friedli, D., Benz, W., 1993, Astron. Astrophys., 268, 65

\bibitem[\protect\citeauthoryear{Friedli}{1994}]{frie1994}
Friedli, D., 1994, in `Mass-Transfer Induced Activity in Galaxies', 
I. Shlosman (ed), Cambridge University Press, p.268 

\bibitem[\protect\citeauthoryear{Froeschl\'{e} et al.}{1997}]{froetal1997}
Froeschl\'{e}, C., Lega, E., Gonczi, R., 1997, Cel. Mech. Dyn. Astron., 67, 
41 

\bibitem[\protect\citeauthoryear{Fux}{2001}]{fux2001} 
Fux, R. 2001, Astron. Astrophys., 373, 511

\bibitem[\protect\citeauthoryear{Gebhardt et al.}{1996}]{gebhardtetal1996}
Gebhardt, K., Richstone, D., Edward, A., Lauer, T.R., Byun, Y.I.,
Kormendy, J., Dressler, A., Faber, S.M., Grillmair, C., Tremaine, S., 
1996, Astron. J., 112, 105

\bibitem[\protect\citeauthoryear{Gebhardt et al.}{2000}]{gerbhardtetal2000}
Gebhardt, K., Richstone, D., Kormendy, J., Lauer, T.R., Ajhar, E.A.,
Bender, R., Dressler, A., Faber, S.M., Grillmair, C., Magorrian, J.,
Tremaine, S., 2000, Astron. J., 119, 1157

\bibitem[\protect\citeauthoryear{Gerhard \& Binney}{1985}]{gerhbinn1985}
Gerhard, O.E., Binney, J., 1985, Mon. Not. R. Astron. Soc., 216, 467

\bibitem[\protect\citeauthoryear{G\"{u}ltekin et al.}{2009a}]{gultetala2009}
G\"{u}ltekin, K., Richstone, D.O., Gebhardt, K., Lauer, T.R., Pinkney, J., 
Aller, M.C., Bender,R., Dressler, A., Faber, S.M., Filippenko, A.V., 
Green, R., Ho Luis C., Kormendy, J., Siopis, C., 2009, Astrophys. J., 
695, 1577

\bibitem[\protect\citeauthoryear{G\"{u}ltekin et al.}{2009b}]{gultetalb2009}
G\"{u}ltekin, K., Richstone, D., Gebhardt, K., Lauer, T.R., Tremaine, S., 
Aller, M.C., Bender, R., Dressler, A., Faber, S.M., Filippenko, A.V., 
Green R., Ho Luis, C., Kormendy, J., Magorrian, J., Pinkney, J., Siopis, 
C., 2009, Astrophys. J., 698,  198 

\bibitem[\protect\citeauthoryear{Hasan et al.}{1993}]{hasetal1993}
Hasan, H., Pfenniger, D., Norman, C., 1993, Astrophys. J., 409, 91  
  
\bibitem[\protect\citeauthoryear{Holley-Bockelmann et al.}{2001}]{holbocketal2001}
Holley-Bockelmann, K., Mihos, J.C., Sigurdsson, S., Hernquist, L., 2001, 
Astrophys. J., 549, 149

\bibitem[\protect\citeauthoryear{Holley-Bockelmann et al.}{2002}]{holbocketal2002}
Holley-Bockelmann, K., Mihos, J.C., Sigurdsson, S., Hernquist, L., Norman, C., 
2002,  Astrophys. J., 567, 817

\bibitem[\protect\citeauthoryear{Jesseit et al.}{2005}]{jessnaabburk2005}
Jesseit, R., Naab, T., Burkert, A., 2005, Mon. Not. R. Astron. Soc., 360,  
1185

\bibitem[\protect\citeauthoryear{Kalapotharakos et al.}{2004}]{kalvogcon2004}
Kalapotharakos, C., Voglis, N., Contopoulos, G., 2004, Astron. Astrophys., 
428, 905

\bibitem[\protect\citeauthoryear{Kalapotharakos \& Voglis}{2005}]{kalvog2005}
Kalapotharakos, C., Voglis, N., 2005,  Cel. Mech. Dyn. Astron., 92,  157

\bibitem[\protect\citeauthoryear{Kalapotharakos}{2008}]{kal2008}
Kalapotharakos, C., 2008, Mon. Not. R. Astron. Soc., 389, 1709

\bibitem[\protect\citeauthoryear{Kandrup \& Sideris}{2002}]{kandrsid2002}
Kandrup, H., Sideris, I., 2002, Astrophys. J., 471, 82

\bibitem[\protect\citeauthoryear{Kaufmann \& Contopoulos}{1996}]{kaufcon1996}
Kaufmann, D.E., Contopoulos, G., 1996, Astron. Astrophys., 309, 381

\bibitem[\protect\citeauthoryear{Kaufmann \& Patsis}{2005}]{kaupat2005} 
Kaufmann, D.E., Patsis, P., 2005, Astrophys. J., 624, 693

\bibitem[\protect\citeauthoryear{Kormendy et al.}{1995}]{kormrich1995}
Kormendy, J., Richstone, D., 1995, Ann. Rev. Astron. Astrophys., 33, 581

\bibitem[\protect\citeauthoryear{Kormendy et al.}{1997}]{kormendy1997}
Kormendy, J., Bender, R., Magorrian, J., Tremaine, S., Gebhardt, K.,
Richstone, D., Dressler, A., Faber, S.M., Grillmair, C., Lauer, T.R.,
1997, Astrophys. J., 482, L139

\bibitem[\protect\citeauthoryear{Kormendy et al.}{1998}]{kormendy1998}
Kormendy, J., Bender, R., Evans, A.S., Richstone, D., 1998, 
Astron. J., 115, 1823

\bibitem[\protect\citeauthoryear{Kormendy \& Ho}{2013}]{kormendy2013}
Kormendy, J., Ho, L.C., 2013, Ann. Rev. Astron. Astrophys., 51, 511

\bibitem[\protect\citeauthoryear{Lauer et al.}{1995}]{Lauer1995}
Lauer, T.R., Ajhar, E.A., Byun, Y.I., Dressler, A., Faber, S.M.,
Grillmair, C., Kormendy, J., Richstone, D., Tremaine, S., 1995,
Astron. J., 110, 2622

\bibitem[\protect\citeauthoryear{McConell \& Ma}{2013}]{mcconma2013}
Merritt, D., 1999, Astr. Soc. Pacific Conf. Ser., 182, 164.

\bibitem[\protect\citeauthoryear{Merritt}{1999}]{mer1999}
Merritt, D., 1999, Astr. Soc. Pacific Conf. Ser., 182, 164.

\bibitem[\protect\citeauthoryear{Merritt }{2013}]{merritt2013}
Merritt, D., 2013, `Dynamics and Evolution of Galactic Nuclei', Princeton 
University Press 

\bibitem[\protect\citeauthoryear{Merritt \& Fridman}{1996}]{merrfrid1996}
Merritt, D., Fridman, T., 1996, Astrophys. J., 460, 136

\bibitem[\protect\citeauthoryear{Merritt \& Quinlan}{1998}]{merrquinl1998}
Merritt, D., Quinlan, D., 1998, Astrophys. J., 498, 625

\bibitem[\protect\citeauthoryear{Merritt \& Valluri}{1996}]{mervalu1996}
Merritt, D., Valluri, M., 1996, Astrophys. J., 471, 82

\bibitem[\protect\citeauthoryear{Merritt \& Valluri}{1999}]{mervalu1999}
Merritt, D., Valluri, M., 1999, Astron. J., 118, 1177

\bibitem[\protect\citeauthoryear{Merritt \& Vasiliev}{2011}]{merrvasil2011}
Merritt, D., Vasiliev, E., 2011, Astrophys. J., 726, 61

\bibitem[\protect\citeauthoryear{Miralda-Escud\'{e} \& 
Schwarzschild}{1989}]{miralda1989}
Miralda-Escud\'{e}, J., and Schwarzschild, M., 1989, Astrophys. J., 339, 752.

\bibitem[\protect\citeauthoryear{Muzzio et al.}{2005}]{muzzcarpwach2005}
Muzzio, J.C., Carpintero, D., Wachlin, F.C., 2005, Cel. Mech. Dyn. Astron., 
91, 173

\bibitem[\protect\citeauthoryear{Muzzio}{2006}]{muzzio2006}
Muzzio, J.C., 2006, Cel. Mech. Dyn. Astron., 96, 85

\bibitem[\protect\citeauthoryear{Norman et al.}{1996}]{normshellhas1996}
Norman, C.A., Sellwood, J.A., Hasan, H., 1996, Astrophys. J., 462, 114 

\bibitem[\protect\citeauthoryear{Patsis et al.}{1997}]{patetal1997} 
Patsis, P.A., Efthymiopoulos, C., Contopoulos, G., Voglis, N. 1997, Astron. 
Astrophys., 326, 493

\bibitem[\protect\citeauthoryear{Pfenniger}{1984}]{pfenn1984}
Pfenniger, D., 1984, Astron. Astrophys., 134, 373

\bibitem[\protect\citeauthoryear{Pfenniger \& de Zeeuw}{1989}]{pfenndezeu1989}
Pfenniger, D., de Zeeuw, T., 1989, in D. Merritt (ed.), `Dynamics of Dense 
Stellar Systems', Cambridge University Press, p.81

\bibitem[\protect\citeauthoryear{Pfenniger \& Friedli}{1991}]{pfefri1991} 
Pfenniger, D., Friedli, D. 1991, Astron. Astrophys., 252, 75

\bibitem[\protect\citeauthoryear{Pichardo et al.}{2004}]{pichetal2004} 
Pichardo, B., Martos, M., Moreno, E., 2004, Astrophys. J., 609, 144

\bibitem[\protect\citeauthoryear{Shen \& Sellwood}{2004}]{shen2004}
Shen, J., Sellwood, J.A., 2004, Astrophys. J., 604, 614

\bibitem[\protect\citeauthoryear{Sparke \& Sellwood}{1987}]{spasel1987} 
Sparke, L.S., Sellwood, J.A., 1987, Mon. Not. R. Astron. Soc., 225, 653

\bibitem[\protect\citeauthoryear{van der Marel et al.}{1997}]{vandermaletal1997}
van der Marel, R.P., de Zeeuw, P.T., Rix, H.W., 1997, Astrophys. J., 
488, 119

\bibitem[\protect\citeauthoryear{Valluri et al.}{2010}]{valur2010}
Valluri, M., Debattista, V.P., Quinn, T., Moore, B., 2010, 
Mon. Not. R. Astron. Soc., 403, 525

\bibitem[\protect\citeauthoryear{Vasiliev \& Athanassoula}{2012}]{vasath2012}
Vasiliev, E., Athanassoula, E., 2012, Mon. Not. R. Astron. Soc., 419, 3268.

\bibitem[\protect\citeauthoryear{Voglis \& Contopoulos}{1994}]{vogcon1994}
Voglis, N., Contopoulos G., 1994, J. Phys. A: Math. Gen., 
27, 4899

\bibitem[\protect\citeauthoryear{Young}{1977}]{young1977}
Young, P.J., 1977, Astrophys. J., 217, 287

\bibitem[\protect\citeauthoryear{Young}{1980}]{young1980}
Young P., J., 1980, Astrophysical Journal, Part 1, 242, 1232-1237  
\end{thebibliography}
\end{document}